\documentstyle[10pt,emulateapj]{article} 
\def\deg{$^{\circ}\,$}
\def\solm{M$_{\odot}\,$}

\def\kms{km~s$^{-1}$}

\def\flux{Jy km s$^{-1}$}
\def\ml{M$_{\rm HI}$/L$_{\rm B}$}
\def\dml{M$_{\rm dyn}$/L$_{\rm B}$\,}

\newenvironment{inlinefigure}{
\def\@captype{figure} 
\noindent\begin{minipage}{0.999\linewidth}\begin{center}}
{\end{center}\end{minipage}\smallskip}

\submitted{Accepted to the Astrophysical Journal}

\begin{document}
\vspace{-7cm}
\title{Galaxy Populations and Evolution in Clusters IV: Deep HI Observations of
Dwarf Ellipticals in the Virgo Cluster}

\author{Christopher J. Conselice$^{1}$, Karen O'Neil$^2$, John S. 
Gallagher$^3$, Rosemary F.G. Wyse$^4$}

\altaffiltext{1}{California Institute of Technology, Mail Code 105-24, Pasadena CA. 91125}
\altaffiltext{1}{NSF Astronomy \& Astrophysics Postdoctoral Fellow}
\altaffiltext{2}{Arecibo Observatory, HC3 Box 53995, Arecibo, PR. 00612}

\altaffiltext{3}{Department of Astronomy, The University of 
Wisconsin-Madison, Madison WI. 53706}

\altaffiltext{4}{Department of Physics and Astronomy, Johns Hopkins
University, Baltimore MD. 21218}

\begin{abstract}

We present in this paper the deepest Arecibo HI observations
of Virgo cluster dwarf ellipticals (dEs) taken to date.  
Based on this data we argue that a significant fraction of 
Virgo cluster dEs recently underwent evolution.  Our
new observations consist of HI 21-cm line observations for 22 classified 
dE galaxies with optical radial velocities consistent with membership in the 
Virgo cluster. Cluster members VCC~390 and VCC~1713 are detected with HI 
masses M$_{\rm HI}$$=~6\times~10^7$~\solm and $8~\times~10^7$~\solm, 
respectively, 
while M$_{\rm HI}$ in the remaining 20 dE galaxies have upper limits as
low as $\sim 5 \times 10^{5}$~M$_{\odot}$. We
combine our results with those for 27 other Virgo cluster dEs with HI
observations in the literature, 7 of which have HI detection claims.  New 
optical images from the WIYN telescope of 5 of these HI-detected dEs, along 
with archival data, suggest that seven of the claimed detections are real,  
yielding a $\approx$15\% detection rate. These  
HI-detected classified dEs are preferentially located near the
periphery of the Virgo cluster.  Three Virgo dEs have observed HI velocity 
widths $> 200$~\kms, possibly indicating the presence of a large dark matter 
content, or transient extended HI.   We discuss the possible 
origins of these objects and argue that they originate from field galaxies 
accreted onto high angular momentum orbits by Virgo in the last few Gyr.  
As a result these galaxies are slowly transformed 
within the cluster by gradual gas stripping processes, associated 
truncation of star formation, and passive fading of stellar populations.  
Low-mass early-type cluster galaxies are therefore currently being produced as 
the product of cluster environmental effects.  We utilize our results in a 
simple model to estimate the recent (past 1-3~Gyr) average mass accretion 
rate into the Virgo cluster, deriving a value of 
$\dot{\rm M}$~$\sim$~50~\solm\,~year$^{-1}$.
  
\end{abstract}

\section{Introduction}                    

Low-mass cluster galaxies (LMCGs), including objects classified as dwarf 
ellipticals (dEs, Ferguson \& Binggeli 1994), dwarf
spheroidals (Gallagher \& Wyse 1994), and irregulars (Gallagher \& Hunter
1984) hold many clues for understanding galaxy formation and 
evolution.  These
are the most common galaxies in the nearby universe (e.g., Ferguson
\& Binggeli 1994), and may be the
first galaxies formed (e.g., Blumenthal et al. 1984; White \& Frenk 1991).  
The relative number densities of dwarfs also increases with local galaxy 
density (e.g., Trentham, Tully, \& Verheijen 2001) with a 
particularly high abundance found in galaxy clusters, including Virgo 
(Binggeli, Tammann \& Sandage 1985; hereafter the Virgo Cluster Catalog or 
VCC).  As such LMCGs offer an outstanding opportunity to understand the 
origin of the lowest mass systems in the universe.

The origins of LMCGs are still a mystery.  Star-forming LMCGs, usually 
classified as dwarf irregulars, are probably recent additions to galaxy 
clusters (e.g., Gallagher \& Hunter 1989; Conselice, Gallagher \& Wyse 2001a, 
hereafter Paper I) as indicated by 
their high gas content, recent star formation, and kinematic characteristics 
(Paper I). Quiescent LMCGs, including dEs\footnote{The term dwarf elliptical 
(dE) is sometimes used in this paper to describe low-mass objects without 
recent star formation,
and with apparently symmetric structures.  Dwarf spheroidals are a
sub-set of dwarf ellipticals with faint magnitudes and low surface
brightnesses (Gallagher \& Wyse 1994).  In our view these two terms are 
interchangeable, and we include both in the general dE designation.
To not bias the interpretation of what Virgo objects with HI detections
are, we use the term low-mass cluster galaxy (LMCG) to describe them, even
though these objects
have been classified by others (e.g,. Binggeli et al. 1985) as dwarf 
ellipticals.}, are low mass 
($<~10^{9}$~\solm), low luminosity (M$_{\rm B}~>~-17$), low surface-brightness 
($\mu_{\rm B}~>~24$~mag~arcsec$^{-2}$) objects with little to no HI gas
(Ferguson \& Binggeli 1994).  Despite their faintness, these 
early-type LMCGs
are more common than their star-forming dIrr counterparts, and likely contain
a mix of stellar population ages and metallicities, and thus possibly 
reflect a variety of origins (e.g., Ferguson \& Binggeli 1994;
Conselice, Gallagher \& Wyse 2002a,b; Papers II \& III). This paper focuses on 
understanding the dE-like LMCGs seen in the Virgo Cluster by searching for HI 
gas that might be left over from their progenitors 
(see Paper III for a detailed discussion of possible formation scenarios).

Due to their low masses and apparently old stellar populations
dwarf elliptical/spheroidal galaxies in the local
universe are outstanding candidates for being among the first galaxies to 
form.   Another 
scenario is that dEs formed after the cluster's initial collapse 
(see Paper I \& III), possibly originating from accreted field galaxies. 
It is however not yet known with certainty (cf. Vigroux et al. 1986; 
Gallagher \& Hunter 1989) if there are any galaxies in nearby clusters
that are currently undergoing morphological transitions from spiral 
$\rightarrow$ dE, 
dIrr $\rightarrow$ dE, or spiral + spiral $\rightarrow$ elliptical, although
inducing these transformations by merging within clusters today is unlikely 
due to the high relative
velocities of cluster members (cf. Conselice et al. 2001b). Dwarf elliptical 
formation is however potentially still ongoing. 

One way to search for objects undergoing formation/evolution into an 
early-type LMCG is to look for galaxies that morphologically
appear as dEs, but have properties suggesting recent evolution from
star-forming, and possibly more massive systems, such
as irregulars or low-mass spirals.  These signs include younger and/or  
metal enriched stellar populations (Paper III) and significant 
atomic gas.    We call these galaxies, containing properties of several 
galaxy types, LMCG transition objects.  In this paper, we present results 
from a survey for HI 21-cm line emission from Virgo early-type LMCGs.
We discover two candidate HI-rich dEs from a sample of 22 observed with the
Arecibo 305-m telescope, which we add to the 7 candidate Virgo Cluster dEs 
with HI detections published in the literature.  
These HI-detected LMCGs have inferred gas fractions that 
are high, placing them in the realm of dwarf irregular or Local Group
transition-type dwarfs, but have morphologies consistent with dwarf 
elliptical objects (Binggeli et al. 1985).  Based on an analysis of these
properties we conclude that seven of these LMCGs are likely transition 
objects, morphologically evolving into early-type dwarfs from
star-forming systems. 

This paper is organized as follows: \S 2 discusses our new observations,
\S 3 presents our basic results after combining our new data with
previously published findings,  \S 4 gives
interpretations of our results, \S 5 gives an estimation of
the current accretion rate into Virgo and \S 6 is a summary.
A Virgo cluster distance of 18~Mpc is assumed throughout this paper, giving
a scale of $\sim$ 5 kpc arcmin$^{-1}$.

\section{Observations and Sample}

The new HI observations we present were taken with the
Arecibo 305 meter radio telescope in April and May 2001.  We observed 22 
Virgo LMCGs classified as dEs by Binggeli et al. (1985) (see Table 1).
This sample of observed objects was chosen from the list of all 
classified dEs in Virgo with known radial velocities, compiled and listed in 
Paper I.   To be observed, an object had to be relatively bright, with 
an apparent magnitude of B~$<$~17 (M$_{\rm B} < -14.0$).   From 
these systems, a random set of 22 objects classified as dEs, or 23\% of the 
total number with known radial velocities covering the range of
Virgo dEs (Paper I), were chosen for 21-cm observations (Table~1). The sample
of observed objects is also spread over the entire angular
extent of the Virgo cluster, as defined in the VCC.

Our 21-cm observations were taken using the Arecibo L-Narrow Gregorian 
receiver. The four Arecibo correlator channels were all centered on the HI 
(21-cm) line, based on the optical determined radial velocities  
(Paper I).  Two different bandwidths were simultaneously observed, with
widths 6.25 MHz and 12.5 MHz,
at two different circular polarizations.  Observations were made
with 9-level sampling, giving each board 2048 channels and providing us
with an unsmoothed resolution of 0.65 and 1.3 km s$^{-1}$ for the two 
different bandwidths, respectively.  Each target was observed using 
standard position-switching 
techniques, with each five minute on-source observation followed by
a five minute `blank sky' observation, which tracked the same
azimuth-zenith angle position of the reflector.  Each on+off pair was 
followed by an observation of a standard noise diode for temperature 
calibration. All data were obtained at night to eliminate solar 
interference.
   
\begin{inlinefigure}
\begin{center}
\resizebox{\textwidth}{!}{\includegraphics{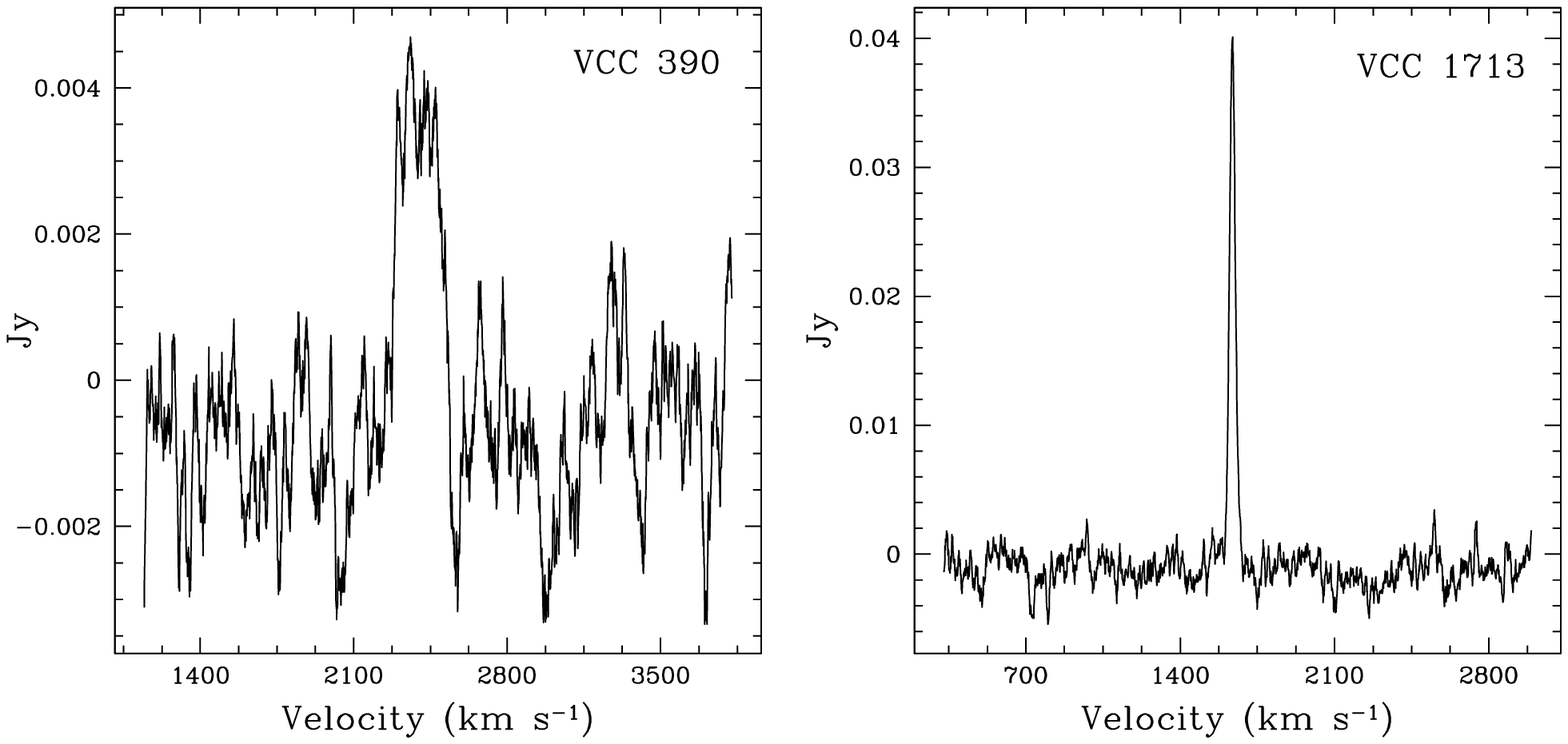}}
\end{center}
\figcaption{Arecibo HI spectra of our two objects with HI detections, 
VCC~390 and VCC~1713.}
\end{inlinefigure}

During each night we also observed strong continuum
sources to check the observatory supplied gain curve.  All calibration
sources were chosen to be non-variable, and to have sizes $<$ 10\% of
the beam width. The flux scales for each source are based on the
calibrations from Baars et al. (1977) and Kuehr et al. (1981).
We also observed galaxies of small angular size with published HI line 
profiles 
from the list in Lewis, Helou \& Salpeter (1985).  These measurements were 
done to confirm the 
accuracy of the internal line flux and frequency calibration.  Table 1 lists
the integration times and 3$\sigma$ detection limits in the HI spectrum
of each object we observed.  On source integration times ranged 
from 0.6~ks to 5.7~ks.

We reduced all of our HI observations using ANALYZ, the Arecibo
data analysis program (O'Neil 2003 in prep).  To obtain fluxes, 
velocity widths and other information for our two new detections, VCC 390 and 
VCC 1713 (Table 2), first order baselines were 
subtracted from the observed spectra.  The baselines were fitted interactively 
and then removed from the HI line profile.  Both single and
double Gaussians were then fit to the HI lines.   
These profiles were fit within the area that occupies the HI emission. That 
is, we start and end the fit where the line distinguishes itself from the 
noise. Since this is ultimately somewhat of a subjective process, we repeated 
it several times to obtain an estimated random error.  This fitting process
gives line widths at 20\% and 50\% of the peak brightness, 
as well as the total flux in each HI line.
 
\section{Results}

\subsection{HI Properties}

Of the 22 galaxies we observed at Arecibo, only 2 have significant HI 
detections (Figure~1). This 
is not an exceptionally low rate, since many previous observations also 
found few, to no, detections of HI in Virgo cluster galaxies 
classified as dEs  (e.g., Huchtmeier \& Richter 1986;  Bothun et al. 1985).
We supplemented our sample by performing a literature search to find other
classified Virgo dEs with HI detections; a further seven objects
with 21-cm detection claims were retrieved.  We discuss these
seven objects in \S 3.5. Table 2 lists all of the Virgo cluster dwarfs 
classified as dEs observed to date at 21-cm.

\begin{inlinefigure}
\begin{center}
\resizebox{\textwidth}{!}{\includegraphics{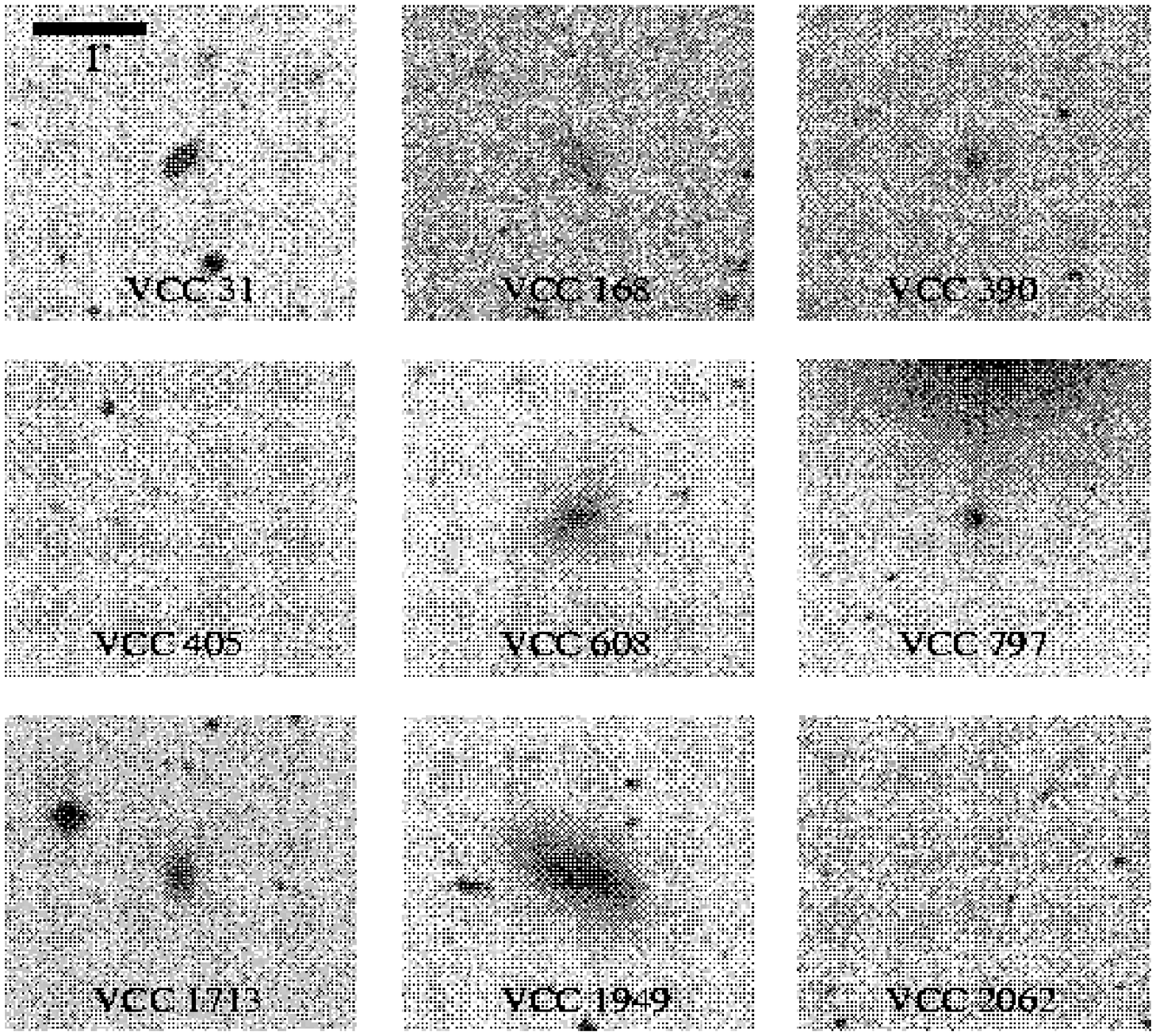}}
\end{center}
\figcaption{Palomar sky survey images centered on the nine candidate
dEs with HI detections.  The size of these images matches the beam of the 
Arecibo telescope at the 21 cm wavelength ($\sim 3\arcmin$).}
\end{inlinefigure}

Since the new observations presented here are the most sensitive ever taken, 
we might expect a higher fraction of 
detections if HI gas was commonly found in early-type LMCGs.  
There are 97 Virgo galaxies classified as dwarf elliptical galaxies with 
measured stellar radial
velocities (Paper I) and 49 (51\%) of these have been observed
in the 21-cm line.  Of these, 9 (19\% of the total) have claimed HI 
detections.  Through the following analysis we conclude that one of these
detections is not significant enough to be considered real (VCC~405)
and we argue that the other is a dwarf irregular galaxy (VCC~2082).  All of 
the published HI detections, including our two new objects, are listed 
in Table 3.

\subsubsection{HI Masses}

We used the flux of the 21-cm line to determine HI gas masses for
each detected LMCG through the formula,

\begin{equation}
{\rm M_{HI} (M_{\odot}\,) } = 2.36 \times 10^{5} D({\rm Mpc})^{2} \int {\rm S}(\nu)\, d\nu,
\end{equation}

\noindent where D is the distance to the galaxy in Mpc, in our case assumed
fixed at 18 Mpc, and
$\int~{\rm S}(\nu)~d\nu$ is the integral of the flux in units of Jy~\kms.
For the undetected galaxies we observed at Arecibo, we set limits on 
HI masses (Table~1) by assuming a 3$\sigma$ detection limit over 60~\kms.  
This provides an average HI upper limit mass of $\sim$8$\times 
10^{6}$~\solm.   
Our deepest observation was for the galaxy VCC~543, where we reach a 
3$\sigma$ limit in the 12.5 MHz band of 0.002~mJy~channel$^{-1}$, 
corresponding to a 3$\sigma$ upper mass limit of 4.6~$\times$~10$^{5}$~\solm.

With these limits we are reaching close to the detectability threshold of 
some Local
Group dwarf ellipticals and spheroidals. For example, the dwarf
ellipticals NGC 205 and NGC 185 have HI masses 3.4$~\times~10^{5}$~\solm
and 1.3~$\times~10^{5}$~\solm respectively (Young \& Lo 1997; Johnson \& 
Gottesman 1993).  Dwarf irregular galaxies such as IC 10 and IC 1613, with 
HI masses of 
1.2~$\times~10^{8}$~\solm and 6.5~$\times~10^{7}$~\solm (Huchmeier 1979; 
Volders \& H\"ogbom 1961), would easily be detected in our survey.  Even
the faintest Local Group dwarf irregular, the Sagittarius DIG, has an
HI gas content of 1.1~$\times~10^{7}$~\solm and would be easily  
detectable at the distance of Virgo in our survey.  

However, a few Local Group
dwarf spheroidals, such as NGC 147 still remain undetected in HI, down to a 
limit
of $\sim$3~$\times~10^{3}$~\solm (Young 1999), and many of the objects we
observe with no detections could have equally low HI gas masses.   Fainter 
dwarf spheroidals, such as Fornax and Leo II, also have extremely low HI mass 
upper limits of $<~10^{4}$~\solm (Young 1999). 

\subsection{Optical Images}

Figure~2 shows
the second generation Digitized Sky Survey (DSS) images for all nine early-type
LMCGs with reported HI detections. This field size was chosen to match the 
Arecibo beam, but detections using other telescopes had different beam sizes. 
In several cases, e.g., VCC 405, nothing obvious
is seen, demonstrating that some objects are very faint.  In most cases,
however, it is clear that these LMCGs are smooth objects with very little
structural
evidence for active star formation, consistent with their classification as
dwarf ellipticals by Binggeli et al. (1985). 

We also acquired WIYN 3.5 meter broad-band Harris R images for five of the nine
Virgo HI detected LMCGs
(VCC~31, VCC~168, VCC~390, VCC~797, VCC~1713).  These shallow images were
taken under non-photometric conditions using the Mini-Mosaic CCD, which has a 
field of view 9.6\arcmin\, $\times$~9.6\arcmin.  Mini-Mosaic consists of 
two SITe 4096
$\times$ 2048 pixel CCDs separated by a gap of 5\arcsec\, and a pixel 
scale of 0.14\arcsec\, pixel$^{-1}$ (see Paper 
II for a further description of 
Mini-Mosaic). The typical exposure times for these five
images were 500 seconds, suitable for obtaining morphological and structural
information.

The five HI detected
LMCGs imaged with WIYN are shown in Figure~3.  With the possible
exception of VCC 1713 these galaxies all appear to be quiescent early-type
LMCGs, confirming the impression from Figure~2.  Figure~3 also shows the 
surface
brightness profiles for these five LMCGs with an arbitrary normalization.
Table~4 lists the quantitative morphological properties for these galaxies, 
including the asymmetry, $A$, (Conselice 1997; Conselice et al. 2000a,b), 
concentration, $C$, (Bershady, Jangren \& Conselice 2000), clumpiness index, 
$S$, (Conselice 2003), as 
well as the S\'{e}rsic profile index and scale radius, $n$ and $r_{0}$, 
according to 
the formula, I~=~I$_{0}$~exp($-$(r/r$_{0}$)$^{1/n}$) (see Paper III 
for details).

The asymmetry parameter, $A$, is a measure of how asymmetric a galaxy
is found by quantifying the residuals after rotating a galaxy's 
image 180\deg by its center and subtracting this new image from the 
original (Conselice et al. 2000a).  Most
early-type galaxies have $A \sim 0$.  The concentration parameter, $C$, is
a measure of how concentrated the light in a galaxy is towards its center.
Most ellipticals have $C > 3.5$, while disks and dwarf galaxies have lower 
values from $2 < C < 3$ (Bershady et al. 2000).  The clumpiness index, $S$, 
is found by quantifying the fraction of spatial high frequency light in a 
galaxy.  
Most early type galaxies have very little high spatial frequency structure,
and thus have clumpiness values $S \sim 0$.  Table~4 is consistent with
the visual impression of these as early-type galaxies.

\begin{inlinefigure}
\begin{center}
\resizebox{\textwidth}{!}{\includegraphics{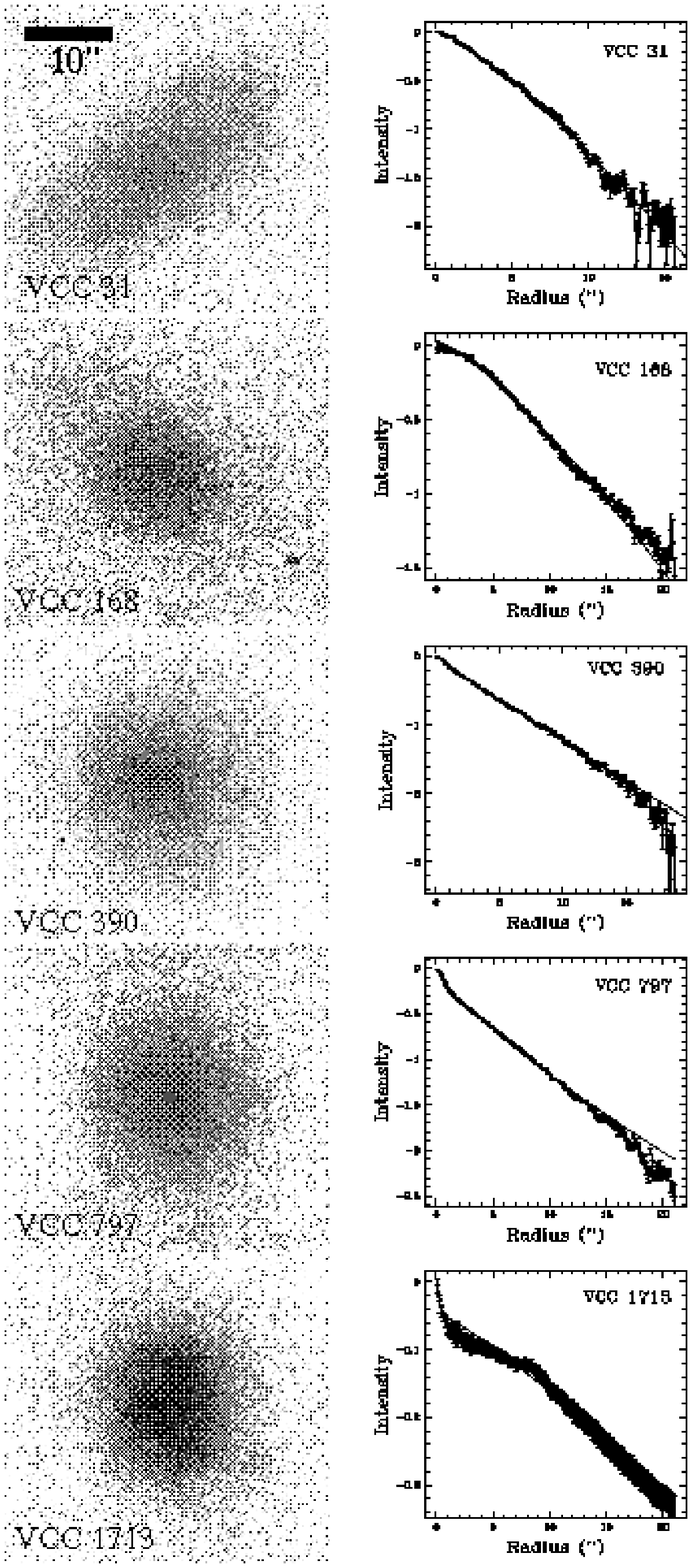}}
\end{center}
\figcaption{Images and surface brightness profiles for the five objects 
detected in HI, imaged with the WIYN 3.5 meter telescope in the R-band.  The 
intensity is log-normalized to zero at the peak intensity and all images
are $\sim 35\arcsec$ on a side, with the scale indicated by the bar on the 
top panel.}
\end{inlinefigure}

\subsection{Dynamical Masses and \dml Ratios}

Although we cannot make any firm dynamical mass claims, we
can compute some estimates based on the likely physical conditions of
each LMCG with an HI detection (Table 3).
A major problem in computing dynamical masses for these galaxies 
is deciding what radius corresponds to the measured
velocity width, and what component of the single gaussian line-widths are due
to rotation, velocity anisotropy, and random turbulent motion.  To estimate
dynamical masses, we assume that the HI velocity width 
represents rotation on orbits that are nearly circular, plus some 
turbulence.    Observations show that the HI sizes of galaxies can be much
bigger than their optical sizes (e.g., Broeils 1992; van Zee, Skillman
\& Salzer 1998), although some
Virgo galaxies have truncated HI profiles (e.g., Haynes \& Giovanelli 1986).
For a size, we use the optical radius estimates for these galaxies given
in the VCC, and listed in Table 3.   We also assume the circular 
velocity (V$_{\rm rot}$) to be half the HI line velocity width at 50\% 
maximum, W$_{50}$, uncorrected for inclination, from our HI spectroscopy.
Normally we would use the W$_{20}$ velocity widths, but these values are
not available for many of the galaxies listed in Table~2.  Using W$_{50}$ 
instead of W$_{20}$ reduces the dynamical
mass, and mass to light ratios, by an average of $\sim$70\%, and increases the 
HI mass fraction by the same amount (Table 3).
We compute lower limits to dynamical masses using the equation:

\begin{equation}
{\rm M_{dyn}} \geq {\rm V_{\rm rot}^{2}} \times {\rm R/G} \sim 0.25 \times {\rm W}_{50}^{2} \times {\rm R}/{\rm G}
\end{equation}

\noindent where we have used the approximation 
V$_{\rm rot} = ({\rm W}_{50}/2)$ with the derived masses listed in Table 3.
The radii we use are from Binggeli et al. (1985) and are also listed
in Table~3.  
From these values we are able to
compute dynamical mass to light ratios, (M$_{\rm dyn}$/M$_{\rm B}$), and gas 
mass fractions, f$_{\rm gas}$~=~M$_{\rm HI}$/M$_{\rm dyn}$ which are 
listed in Table 3.   Several
of these galaxies have very high mass to light ratios as discussed below, 
while others have rather modest values with \dml $\sim$ 1 in solar units.  
The highest \dml values are 
found for VCC 405 and VCC 2062, which we conclude in \S 3.5 are likely not
dwarf ellipticals or real detections.   We also discuss alternative 
processes that can potentially 
create large HI velocity widths besides deep gravitational potential wells.

\subsection{New Detections}

In this section we discuss the 21-cm and optical properties 
for each of our new Arecibo HI detections and give some possible
interpretations of their measurements.  We wish to stress that both of
these detections have only occured with Arecibo and its large
beam, and confusion with other galaxies could be an issue (Hibbard
\& Sansom 2003).

\subsubsection{VCC 390}

In the DSS and WIYN images (Figures 2 and 3) 
VCC 390 appears as a small compact galaxy.  Using the WIYN image of this
galaxy (Figure~3) its scale length is fit as 0.29$\pm 0.003$ kpc (Table~4).
This object is classified as a dE3 in the VCC and has an absolute magnitude of
M$_{\rm B} = -14.4$, has low asymmetry, low concentration, and low 
clumpiness values (Table~4), all consistent with it being
a dwarf elliptical (Conselice 2003).  It is also fairly well fit by
an exponential profile with a S\'{e}rsic index, n = 0.95$\pm 0.01$.  This 
galaxy is however fainter towards its outer parts than
its S\'{e}rsic fit.

The observed wavelength of the HI line for VCC 390 gives a heliocentric 
radial velocity of 
2400$\pm$10~\kms\, (Figure~1), in relatively good agreement with its optical 
radial velocity, 2479$\pm$38~\kms\, computed though cross-correlation fits of
absorption features in the optical VCC~390 spectrum shown in Figure~4.
This is a WIYN Hydra Multi-Object Spectrograph (MOS) spectrum taken in 
April 1999 as part of a radial velocity survey (Paper I).
This is a typical early-type galaxy spectrum showing absorption features
such as the Mgb triplet at 5175 \AA.  No emission lines are seen; these
would be an indication of active star formation, as seen in Virgo
dwarf irregular galaxies (e.g., Gallagher \& Hunter 1989; 
Heller et al. 1999).  The spectrum of VCC~390 is also similar to spectra
of other Virgo dwarf ellipticals (see Paper I).

\begin{inlinefigure}
\begin{center}
\resizebox{\textwidth}{!}{\includegraphics{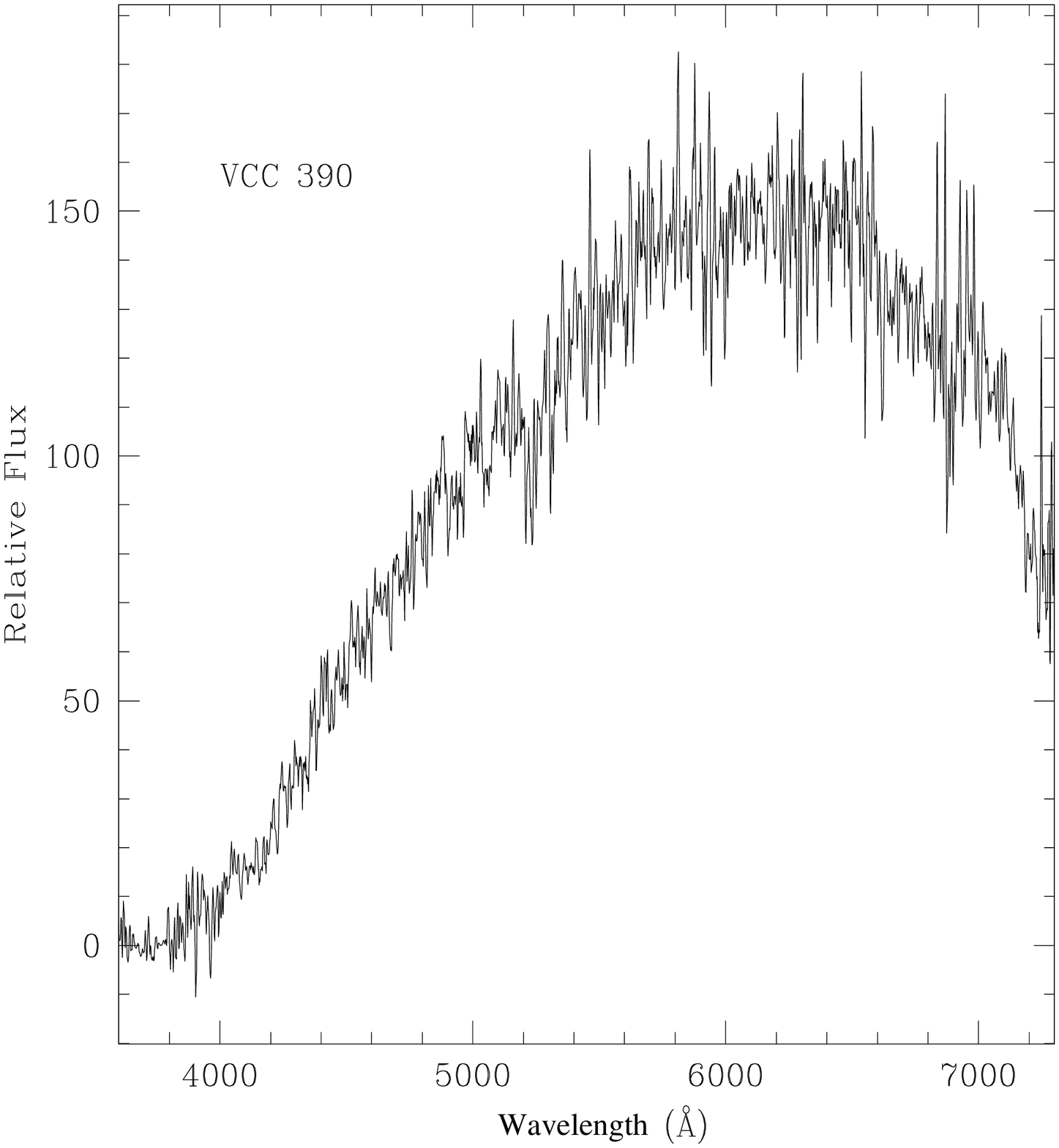}}
\end{center}
\vspace{-0.3in}
\figcaption{Optical spectrum of the dwarf elliptical galaxy VCC~390, taken
with the WIYN 3.5m telescope with an arbitrary flux normalization.}
\end{inlinefigure}

The HI line of VCC 390 presents an interesting puzzle due to its large
velocity width (Table 3).   The 20\% and 50\% FWHM velocities are
286$\pm 5$ \kms\, and 285$\pm 7$ \kms\, with a total 
integrated HI flux of 0.8$\pm 0.01$ \flux.  It is only one of three galaxies 
in our sample that has a velocity width $> 200$ \kms.  Due to its faint 
magnitude, it also
has an inferred high dynamical mass to light ratio (M$_{\rm dyn}$/L$_{\rm B}$) 
of $\sim$45 in solar units (Table 3).  However, it is possible that some of
this line emission originates from contamination in the side-lobes
due to the nearby galaxy NGC~4277, which is at a velocity 2504 \kms\,, and has 
an HI line profile that overlaps in velocity with VCC~390's (see van Driel et 
al. 2000). We investigated this possibility by overlaying the Arecibo beam 
map onto a DSS image of VCC~390 and NGC 4277.  We 
found that the outer lobes of the profile as centered on VCC~390 slightly 
overlapped with NGC~4277.  Short of this, the large velocity width could 
be the result of gas stripping.  
It is however impossible to determine using our current
data how much these sources are contributing to the VCC~390 signal.  
Understanding this will require mapping the spatial 
extent of the HI distribution of VCC~390 and comparing it with that of
NGC~4277.

\subsubsection{VCC 1713}

The other low-mass galaxy with a new HI detection is VCC 1713. 
Compared to VCC 390, VCC 1713 has a narrower velocity HI profile, 
with $W_{50}$ = 46$\pm 8$ \kms\, and $W_{20}$ = 60$\pm 6$ \kms, more commonly 
expected for a low-mass
galaxy.  The total HI flux is slightly higher than that
of VCC 390, with an integrated value of 1.1 \flux, giving a total 
HI gas mass of 8 $\times$ 10$^{7}$ \solm. Despite its narrow HI line
width, this galaxy has a rather bright absolute magnitude of 
M$_{\rm B} = -16.2$. This gives a \dml value of $>$ 0.4 in solar
units. 

It is also listed as an 
uncertain member of the Virgo cluster in the VCC; however, its velocity of 
1655~\kms\, (Grogin et al. 1998) is consistent with cluster membership.
The morphological appearance (Figure 2 \& 3) and quantitative morphology of 
this galaxy suggests that it is a
dwarf elliptical-like object (Table~4), with low asymmetry and
concentration values.  The clumpiness index $S$ is 
moderately high suggesting the presence of some low level of star formation.  
There are however no measured colors for VCC~1713, which if it were very blue
would qualify as a dwarf irregular object.
This object is our best candidate for being an object in an ongoing early
transition phase between a star-forming and quiescent LMCG.

\subsection{Previous Detections}

The seven following objects, all classified as dEs, were previously detected 
in HI.  In some cases we also discuss the
possibility that some of these objects are not early-type LMCGs, but are 
misclassified star-forming galaxies, or spurious HI detections.

\subsubsection{VCC 31}

VCC 31 has a magnitude of M$_{\rm B} = - 16.4$ and was detected in HI using 
the Arecibo telescope by van Zee, Haynes \& Giovanneli (1995)
and at Nan\c{c}ay by van Driel et al. (2000).  Van Zee et al. (1995) determined
that this galaxy does not have an extended HI distribution, and has
a M$_{\rm HI}$/L$_{\rm B}$ value high enough to be considered a compact
dwarf.    Van Driel et al. (2000) also detected this object in HI, although
they find a much higher $W_{20}$ value (312 \kms) than van Zee et al. (1995)
(132 \kms).   This indicates that VCC~31 may have 
extended gas, as the velocity width and flux of the Nan\c{c}ay observations
are much larger than the Arecibo results, which used a smaller 3.5\arcmin\, 
beam.  There are also no nearby galaxies that would have confused the 
Nan\c{c}ay results, further suggesting that there is extended HI
gas around VCC~31.
This galaxy appears symmetric and dwarf elliptical-like
on the DSS and WIYN images (Figure 2
and 3). Its quantitative morphology (Table~4) is also consistent with a
dwarf elliptical classification.  
Using the van Zee et al. (1995) parameters of the HI line for this galaxy,
we find a \dml ratio $\sim$ 2.4 in solar units, while the van Driel et al.
(2000) values give ratios a factor of $\sim 5$ higher.

\subsubsection{VCC 168}

VCC 168 was first detected in HI using Arecibo by Hoffman et al. (1987), who 
classified
this object as a dE2 or ImIV based on the VCC.  This object was later 
classified by Binggeli \& Cameron (1993) as a true dE2.  It is not 
visible on the DSS images (Figure 2), but clearly appears as a dwarf 
elliptical-like object on the WIYN image, with a quantitative morphology
consistent with this interpretation (Table~4) including a light profile
well fit by an exponential (Table~4; Figure~3). 
Heller et al. (1999) searched for H$\alpha$ in VCC 168, but found none.  
VCC 168 is also symmetric, with
little internal structure, based on its image in Heller et al. (1999) and
in Figure~3, implying that 
the dwarf elliptical classification is the correct 
one.\footnote{Heller et al. only find five objects out of 43 without 
H$\alpha$ emission in their survey of Virgo cluster irregulars, one of 
which is VCC 168.  This further suggests that this object is not a true
irregular. The irregular classification of VCC 168 by Heller et al. is solely
based on its HI detection by Hoffman et al. (1987).}   Almoznino \& Brosch 
(1998) observed VCC 168 in BVR, finding
a color $(B-V)$ = 0.76$\pm$0.17, the reddest $(B-V)$ color in their sample.
This suggests an older stellar population, $> 2 $ Gyr since the
cessation of significant star formation. This is the case even if the stars are
metal rich (Worthey 1994); VCC 168 thus appears to be a 
quiescent LMCG with gas. Its HI mass of $3 \times 10^{7}$ \solm
is unusually high for such a faint dwarf elliptical.  It also has an HI
line width of W$_{20}$ = 64 \kms\, suggesting a dynamical mass to light
ratio of \dml $\sim$ 2.5 (Table 3).

\subsubsection{VCC 405}

VCC 405 is the faintest object in our HI detected sample with B~$\sim$~20 
(VCC) or M$_{\rm B}~\sim -11.5$, and it does not 
appear on DSS images (Figure 2).   This object was marginally
detected with Arecibo at 21-cm (Duprie \& Schneider 1996) with a single
observation, and thus the detection could be
spurious. Based on their HI spectrum and the very unlikely physical 
parameters derived from
its HI profile (Table 3), we conclude that this object is indeed 
probably a false HI detection, and do not consider it in our subsequent 
analysis.

\subsubsection{VCC 608}

VCC 608 (NGC 4323), classified as a dE4, N in the VCC, is the second brightest 
Virgo dwarf elliptical detected in neutral
hydrogen, with M$_{\rm B} = -16.6$.  It was first detected in HI
by Huchtmeier \& Richter (1986). VCC~608 also has a fairly red color with 
$(B-V)$ = 0.87$\pm$0.06 (de Vaucouleurs et al. 1991), consistent with an old
stellar population (see the argument in \S 3.5.2).  Ryden et al. (1999) 
determined that this galaxy has a mean ellipticity of 0.39 with
isophotal fits giving $<$a$_{4}$/a$>$ values consistent with a disky 
structure. Gavazzi et al. (2001) fit surface photometry
in the near-IR H(1.6$\mu$) band, finding a good exponential fit, and a low 
concentration of light, further confirmation that this is morphologically 
similar to a dwarf galaxy.  VCC~608 also has a
\dml ratio $\sim$ 2.5, based on its HI line width.  It also contains a high 
HI gas content of $\sim$ 10$^{9}$ \solm, and has one of the highest HI gas 
mass fractions in our sample (Table 3).

\subsubsection{VCC 797}

VCC 797 with $W_{50} =$ 28 \kms\, has the narrowest HI line of any galaxy in
our sample, and the smallest optical
scale length of 0.24 kpc.  The internal velocity dispersion is low, 
indicating that its observed HI velocity width may be dominated by 
turbulence.  
It is also the only low-mass galaxy that is near a giant galaxy, in this
case about 3\arcmin\, south of M85 (Figure~2). HI in VCC 797 was discovered 
by Burstein, Krumm \& Salpeter (1987), who detected it during
21-cm observations of M85.  Due to the presence of HI, Burstein et al. 
(1987) considered VCC~797 to be a dwarf irregular,
although it is classified as a dE3,N in the VCC.  The VCC classification
is based on higher resolution
images than those available to Burstein et al. (1987). Burstein et al.
(1987) however only marginally detect the object at a 4$\sigma$ level of
significance.  We tentatively consider this to be a HI detected dE, and not a 
spurious signal, although this should be confirmed by observing it again at 
21-cm with higher sensitivity.

\subsubsection{VCC 1949}

VCC 1949 is the brightest galaxy (M$_{\rm B} = -17.1$) in our sample, and it 
appears as an early-type dwarf on DSS images (Figure~2).  It
was classified as a dSBO(4),N or dE(6,4),N in the VCC\footnote{Bingelli
\& Cameron (1993) list VCC 1949 as a dS0.}.  It was
first detected in HI by Huchtmeier \& Richter (1986), and later
re-observed at 21-cm by Haynes \& Giovanelli (1986).  The optical image of  
this galaxy is well fit by an exponential profile (Binggeli \& Cameron 1993).
VCC 1949 is one of three galaxies in the present sample with very large 
($ > 250$ \kms) inferred HI
velocity widths (Table 3), giving it a high measured dynamical mass to light 
ratio of \dml $\sim$ 20 in solar units.  There are no obvious companions which
could produce this large line width from confusion.

\subsubsection{VCC 2062}

VCC 2062 has been studied previously in HI by both Hoffman et al. (1993)
and van Driel \& van Woerden (1989).  It is about 2 kpc (0.4 arcmin) away 
from the irregular galaxy NGC~4694, with which it may be interacting.  VCC~2062
was classified as a dE in the VCC, although the DSS image in Hoffman et al.
(1993) suggests that this classification is incorrect, and that the galaxy
is more likely an irregular.  Hoffmann et al. also found
very blue colors for this object, with $(B-V)$ = 0.35 and $(V-R)$ = 0.19.
These colors indicate that star formation has 
recently occurred, as an extremely  metal poor stellar population which can 
also produce blue colors, would have to be less than 700 Myrs old
to produce these colors (Worthey 1994). The presence of young
stars reveals that this object is probably not a classical dE (i.e., old 
stars and no gas), despite its classification in the VCC.

Due to its fairly high M$_{\rm HI}$/L$_{\rm B}$ ratio of $\sim$ 55 this 
object is also by far the most gas rich member of our sample.  The 
high \ml\, ratio exceeds that of typical values found for blue compact dwarf 
galaxies (van Zee et al. 1995) and irregular galaxies (e.g,. Roberts \& 
Haynes 1994), another indication that it may be an interacting system.

In summary, we conclude that 7 of the 9 detections of HI in LMCGs are 
genuine, and reject as spurious the HI dE detection claims for VCC~405 and 
VCC~2062.

\section{Interpretations}

\subsection{Incorrect Morphological Identifications}

Although the galaxies discussed in this paper are classified as dEs 
in the VCC, other studies have given different classifications for a few of
these objects, mostly the dwarf irregular class.  Could
all of these HI detected LMCGs in our sample be misclassified
dwarf irregulars?  To some degree the problem is
semantic assuming that the VCC used a consistent classification approach 
throughout the Virgo cluster.  The identification in the VCC of
a dE therefore may be biased, but as long as this bias is consistently 
applied throughout, then the VCC classified `dEs' should represent a 
homogeneous population equally distributed across the cluster. 

\begin{inlinefigure}
\begin{center}
\resizebox{\textwidth}{!}{\includegraphics{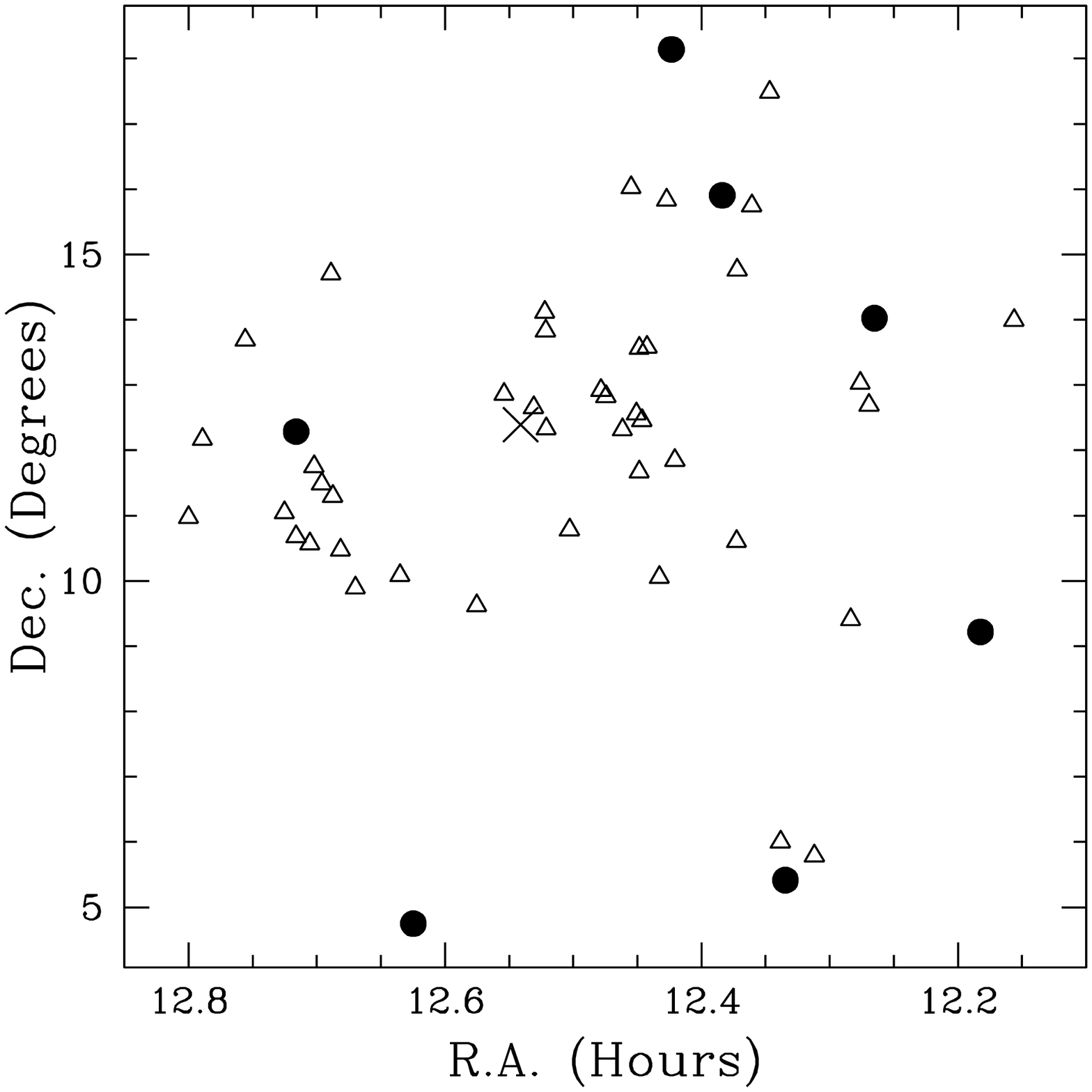}}
\end{center}
\vspace{0.3in}
\figcaption{The distribution on the sky of dEs observed at 21-cm in the Virgo 
cluster.
The open triangles represent non-detections while the solid circles
are the location of the dEs with detected HI.  The cross toward the
center is the location of the giant elliptical M87.}
\end{inlinefigure}

We also reject the incorrect morphological identification explanation for
the following reasons.  The qualitative and quantitative morphologies
for these objects, as seen in the DSS frames and in the WIYN images,
suggest quiescent systems, with no obvious 
evidence for star formation, with the possible exception of VCC~1713. The  
available colors for VCC 168 (\S 3.5.2) and VCC 608 (\S 3.5.4) also 
demonstrate that some HI detected early-type LMCGs 
potentially have older dE-like stellar populations.
The HI gas contents for these systems are also too low for these 
objects to be misclassified normal dwarf irregulars, as Virgo dIrrs
have higher HI gas masses and fluxes than those listed in 
Table~3 (e.g., Hoffman et al. 1987; Gallagher \& Hunter 1989; see also
Table~5).  Therefore we conclude that these HI detected LMCGs are probably not 
misclassified dwarf irregulars, with the one exception being VCC~2062 which
is not included in the following analyses.

\subsection{Patterns of HI Depletion and Orbital Structure}

In Figures~5 and 6 we show the spatial pattern of HI depletion of Virgo
classified
dwarf elliptical galaxies,  where in Figure~5 galaxies with HI detections 
are plotted as large filled
circles and non HI-detections as open triangles.  All of the LMCGs classified 
as dEs with HI detections are projected towards
the outer parts of the cluster.  This is quantitatively shown in
Figure~6, which plots the fraction of LMCGs classified as dEs with HI
detections as a function of projected
distance from the luminosity weighted center of the Virgo cluster, as defined
by Sandage et al. (1985). 
Within a projected distance of 0.5 Mpc of the center, the 
lower limit to the true 3-dimensional distance, we find no galaxies 
classified as dEs with HI detections.  

\begin{inlinefigure}
\begin{center}
\resizebox{\textwidth}{!}{\includegraphics{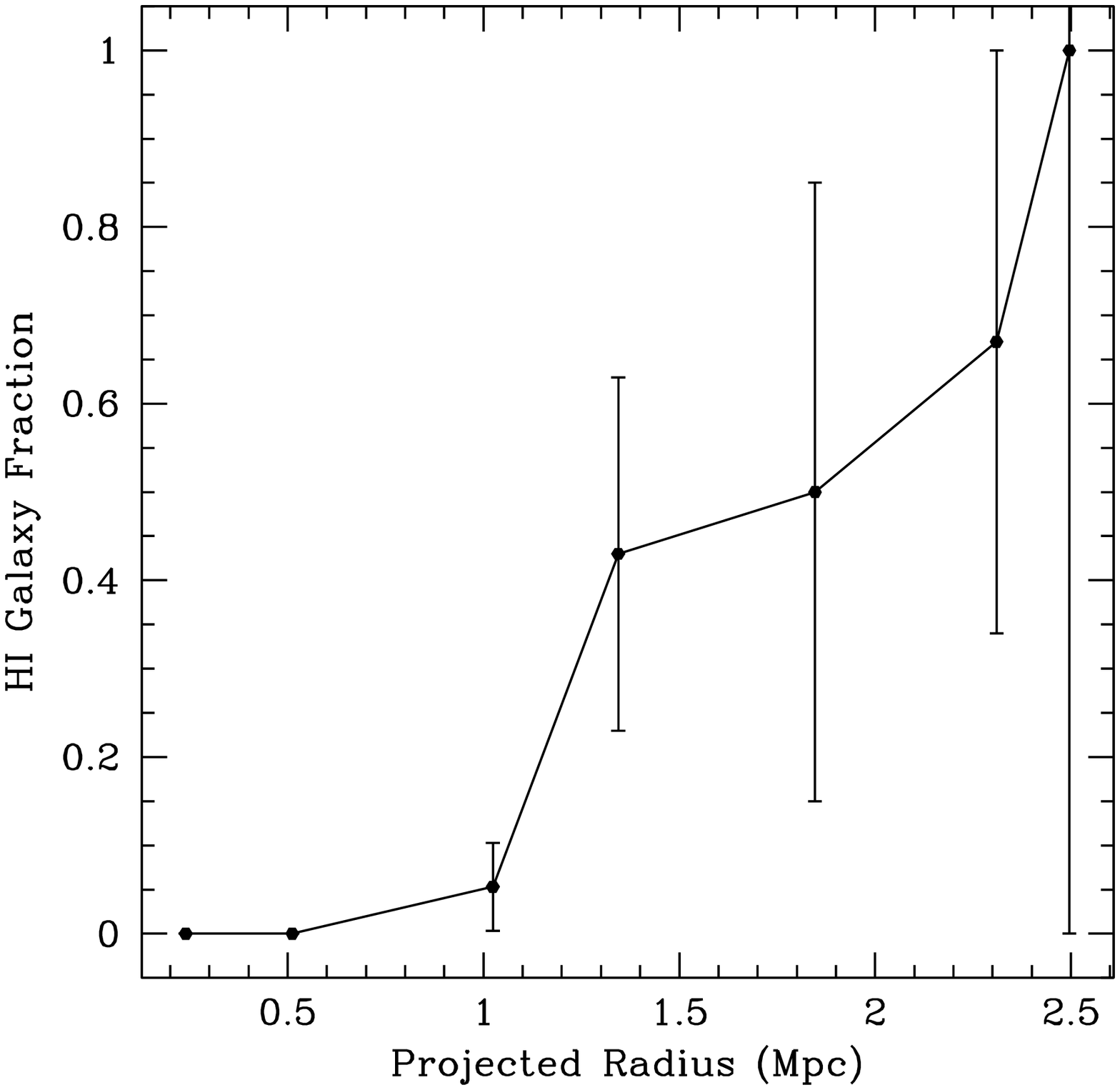}}
\end{center}
\figcaption{The fraction of Virgo classified dwarf elliptical galaxies 
detected in HI as a function of projected distance from the center of the 
cluster.}
\end{inlinefigure}

The presence of most of the HI LMCGs in the outer parts of Virgo may help 
reveal the origin of these objects.  One obvious explanation is that these 
galaxies have just been accreted into the cluster and have not yet crossed the
core.  The other possibility is that these galaxies have orbits that
take them through the core of the cluster and somehow retain or replenish 
their HI gas.  For the purposes of argument we discuss the infall case, and 
two idealized orbiting scenarios: radial and high angular momentum orbits. 

Cosmological models of galaxy clusters show that most cluster members 
are originally on moderate, or low, angular momentum orbits (e.g., Crone, 
Evrard, \& Richstone 1994; Ghigna et al. 1998; Huss, Jain, \& Steinmetz 
1999).  Gas stripping processes are also most effective for galaxies on 
nearly radial orbits (Vollmer et al. 2001). The survival of HI in low mass
cluster galaxies therefore depends on orbital parameters, such that galaxies in
more circular orbits that avoid the core are more likely to retain their 
HI. Signatures of this effect also may be seen in samples of giant galaxies 
(e.g., Dressler 1986; Solanes et al. 2001) and might also be
present in some LMCGs. However, Solanes et al. do not find clear
evidence for this behavior in Virgo, possibly as a result of complications
associated with interactions with substructure in this cluster (see also 
Vollmer et al. 2001).

Evidence that our HI LMCGs are on high angular momentum orbits includes the 
distribution of their radial velocities. The expected signature for
galaxies on high angular momentum orbits would be a flat-topped distribution in
velocities.  Galaxies on radial orbits would have a gaussian distribution
peaked at the mean velocity of the cluster.   The average velocity of the HI
LMCGs is relatively large and not centered at the average velocity of
the cluster, and the distribution of velocities is non-uniform, different 
from expectations of both the
radial and high angular momentum distribution.  Although our distribution
is not flat-topped, it is clearly not uniformly distributed about the mean
cluster velocity.   The velocity distribution of the HI rich early-type 
LMCGs velocity distribution is shown in Figure~7, where
the shaded part displays the distribution of the HI detected LMCGs, and the
open histograms are the velocity distribution of the non-detections.  Although
we are in the regime of small number statistics, we can make
some arguments based on the data available.  
 
\begin{inlinefigure}
\begin{center}
\resizebox{\textwidth}{!}{\includegraphics{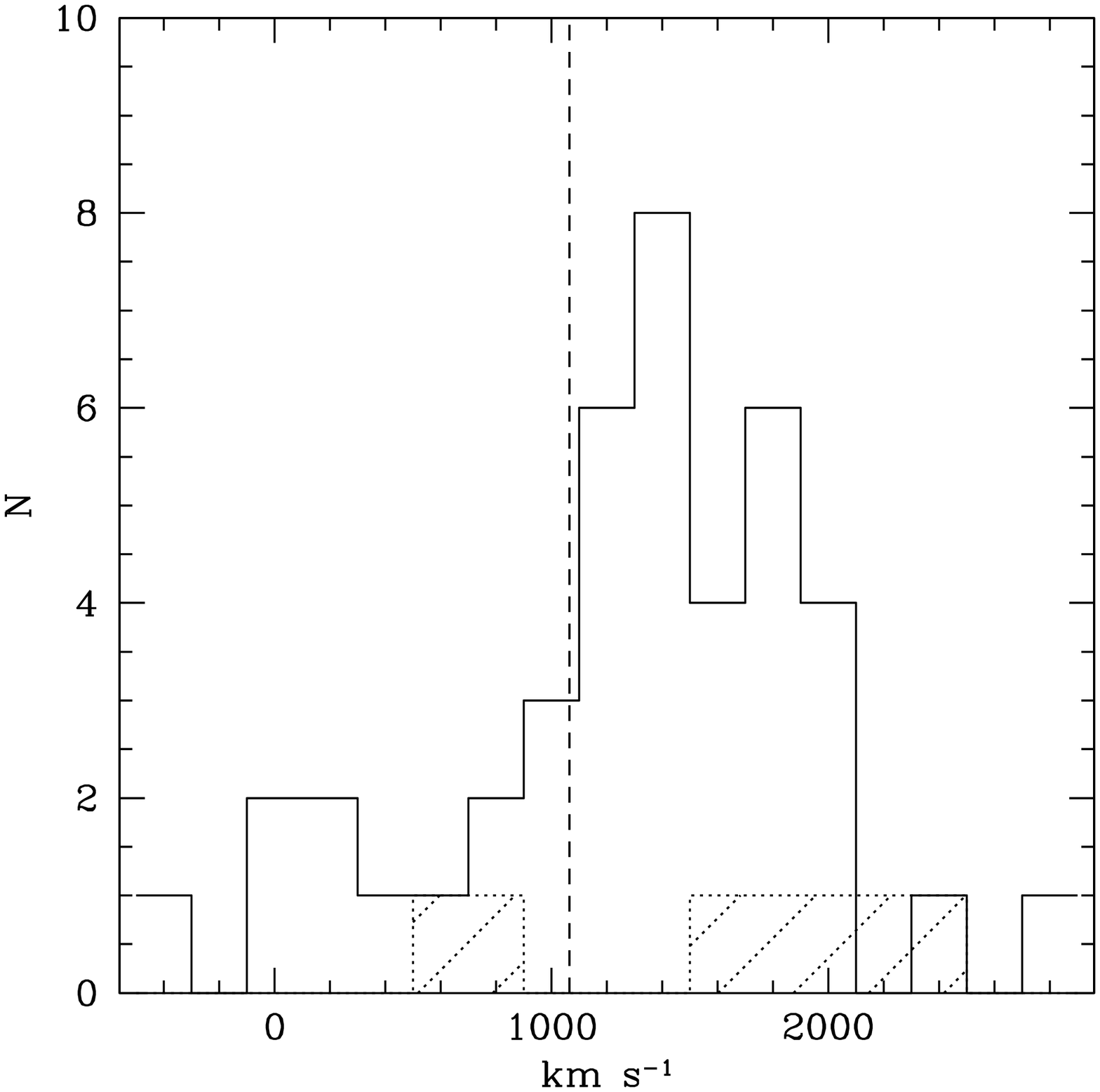}}
\end{center}
\figcaption{Histogram of radial velocities for dEs searched for in HI. The open
histograms represent the dE non-detections, while the shaded histogram
is for Virgo dEs with HI detections.  The vertical dashed line shows the
mean velocity of the Virgo cluster, from Paper I.}
\end{inlinefigure}

The average  velocity of the non HI-detected objects is 1287$\pm$183~\kms, 
and the  HI-detected objects have an average velocity of 1669$\pm$553 \kms.  
The velocity dispersions are very similar with $\sigma$ = 685~\kms\, and
697 \kms\, for the non-detected and detected objects, respectively. 
However, for galaxies at $> 1.5$ Mpc projected separation from the center of 
Virgo, the velocity dispersion for the HI LMCGs is 200 \kms\, higher than for
the LMCGs with non HI detections.  A Kolmogorov-Smirnov test
reveals a 24\% probability that these two distributions are similar.
Figure~8 shows the histogram of 
magnitudes for the HI detected (shaded) and non-detected (open) objects,
demonstrating that unlike the radial velocities, the magnitudes of the
detected objects are not significantly skewed towards any particular values, 
although they are on average slightly fainter.  

\begin{inlinefigure}
\begin{center}
\resizebox{\textwidth}{!}{\includegraphics{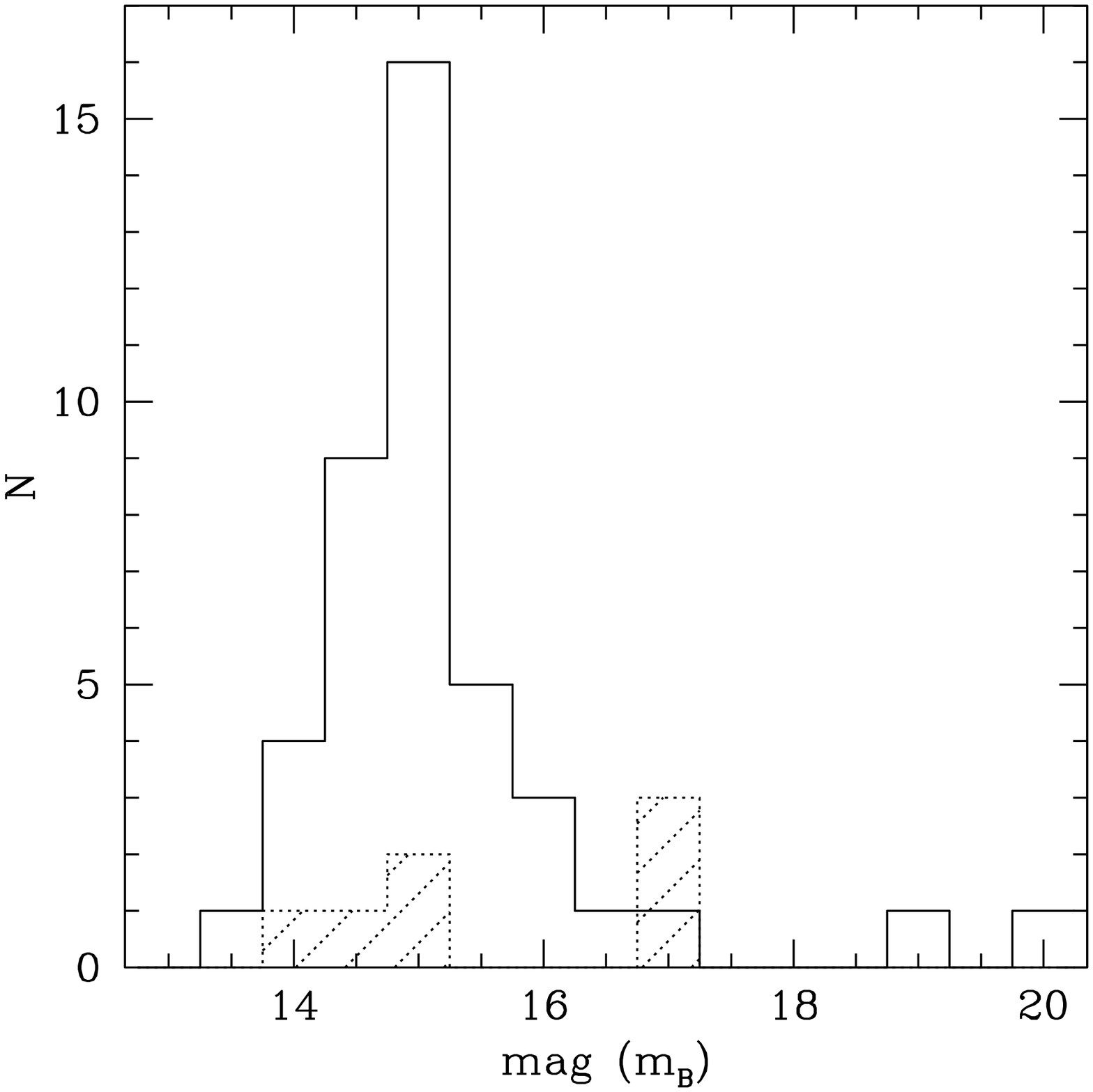}}
\end{center}
\figcaption{Histograms of apparent B-band magnitudes for non-detected (open) 
and HI detected (shaded) Virgo cluster dEs.  The shadings are the same as in 
Figure 7.}
\end{inlinefigure}

To determine the significance of the $\sim 400$ \kms\, difference in average
velocity between the HI detected and non-detected LMCGs we perform a
Monte Carlo simulation based on the velocities of dEs in Virgo (Paper I).
Using the $\sigma$ and the mean velocity of dEs from Paper I, we randomly
select seven galaxy velocities from this Gaussian distribution, and compute
the mean of these seven values.  This exercise reveals
less than a 1\% chance ($\sim 3 \sigma$) that the $400$~\kms\, difference
is random.  Therefore the fact that the velocities of
these HI detected LMCGs are not near the mean cluster velocity of
1064$\pm$34 \kms\,, and are found in the outer parts of Virgo, is an 
indication that a fraction of these HI LMCGs are not on radial 
orbits and are likely on high angular momentum ones with some velocity
substructure induced perhaps by galaxies infalling within groups.  

\subsection{Gas Stripping, Replenishment and Star Formation}

The spatial pattern of LMCGs with detectable HI suggests that gas stripping 
has occurred through interactions
with the intracluster medium (Dressler 1986; Vollmer et al. 2001).  
Such stripping can happen in a number of ways.  
The most commonly suggested method is ram-pressure stripping, which can 
deplete a dwarf rapidly in the inner regions of a cluster, in typically less
than 0.1 Gyr (Mori \& Burkert 2000) for a CDM like halo, 
although this depends on
the time for shock waves to traverse the galaxy, the form of the 
galaxy's gravitational potential, and properties of the ICM.
A more gradual mass loss may also occur due to Kelvin-Helmholtz instabilities
(Nulsen 1982), although our estimate below suggests that it is not likely
important for Virgo cluster dwarfs.  

\newpage
\subsubsection{Ram-Pressure Stripping}

The ram-pressure on a galaxy from the interaction of intracluster gas   
is given by the equation P~=~$\rho_{\rm ICM} v_{\bot}^{2}$, where 
$v_{\bot}$ is the velocity perpendicular to the plane of a galaxy.
The efficiency of gas stripping also depends upon the 
gravitational potential $\phi$ of a galaxy, i.e., its ability to hold onto 
its material.  The vertical gravitational acceleration
of ISM clouds in a disk galaxy is given by $\partial \phi$ / $\partial z$.
If we define a galaxy's gas surface density as $\sigma$, then ram-pressure 
stripping occurs in the galaxy when

\begin{equation}
\frac{\partial \phi}{\partial z} \sigma < \rho_{\rm ICM} v_{\bot}^{2}.
\end{equation}

\noindent Here $\rho_{\rm ICM}$ is the density of the intracluster
medium of the Virgo cluster, and $v_{\bot}$, as before, is the relative 
velocity 
perpendicular to the plane of the disk.  This formula, first proposed by
Gunn \& Gott (1972), is found to be a good approximation based on detailed
simulations (e.g., Abadi, Moore, \& Bower 1999; Vollmer et al. 2001). 

To determine the intracluster medium density, $n(r)$, as a function of 
radius, we use a $\beta$ model (Cavaliere \& Fusco-Femiano 1976), and the 
parameters for the Virgo cluster given by Vollmer et al. (2001),

\begin{equation}
n(r) = n_{0}\left(1 + \frac{r^{2}}{r_{0}^{2}}\right)^{-(3/2)\beta}.
\end{equation}

\begin{inlinefigure}
\begin{center}
\resizebox{\textwidth}{!}{\includegraphics{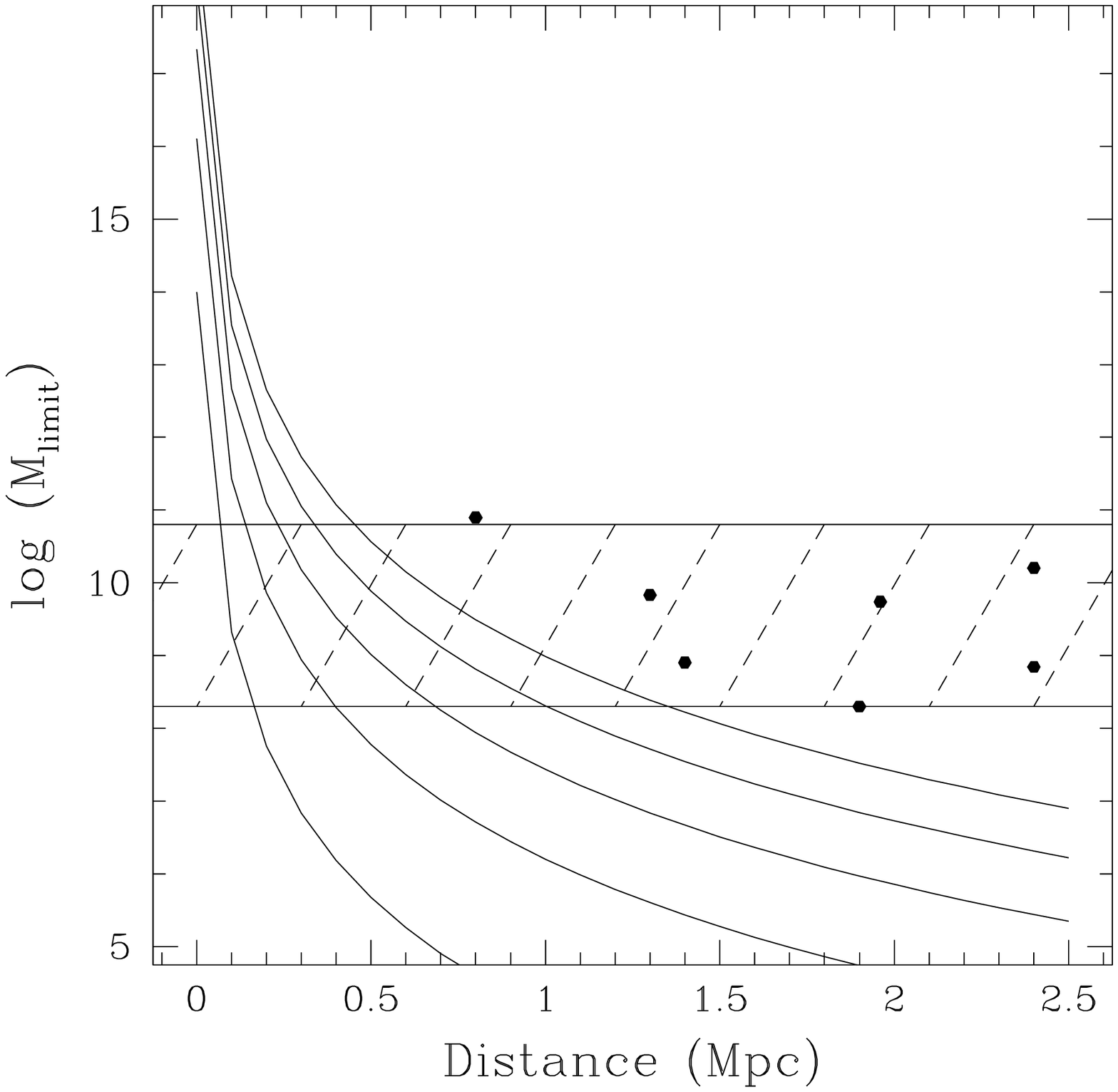}}
\end{center}
\figcaption{Relationship between the minimum galaxy mass, M$_{\rm limit}$, 
above which HI gas is expected to survive ram-pressure stripping, as
a function of distance from the Virgo cluster center.  The five solid lines 
show this relationship
for galaxy orbital velocities relative to the ICM of $v =$ 250, 500, 750, 
1000 and 1250 \kms\, at increasing mass limits.  The solid points show the
projected locations of the seven HI LMCGs studied in this paper which are all
above the minimum mass limits.  The hatched region shows the range of total
masses for our HI LMCGs, and dwarf galaxies in general.}
\end{inlinefigure}

\noindent In this equation $r$ is the physical 
distance from the center of the cluster, $n_{0}$ is the central ICM gas 
density, which is $n_{0} = 4 \times 10^{-2}$ cm$^{-3}$, and $r_{0} = 13.4$
kpc, and $\beta = 0.5$ for Virgo intracluser gas (Vollmer et al. 2001).  
By knowing the gas
density as a function of radius, the fraction of a galaxy's mass in the gas
phase, and the velocity of the gas with respect to the intracluster medium, 
the minimum mass, M$_{\rm limit}$, a galaxy must have to avoid being
stripped of gas by ram-pressure can be computed analytically (Mori \& Burkert 
2000) for CDM halos.
If we assume a gas mass fraction of $\sim 0.1 - 0.3$ for a typical LMCG, we 
can compute M$_{\rm limit}$ as a function of $v_{\bot}$ and $n(r)$.  Using 
eq. (4) 
for the gas density as a function of radius in Virgo, we plot in Figure~9 the
mass limit M$_{\rm limit}$ as a function of radius at velocities $v_{\bot}$ 
from  250 \kms\, to 1250 \kms.  Also plotted
on this diagram, as solid points, are the dynamical masses and projected 
distances from the cluster center for each of the seven LMCGs with confident 
HI detections.  This figure shows that at
the relatively large distances of the detected HI LMCGs from the Virgo center, 
their masses are high enough such
that ram-pressure stripping cannot rapidly deplete them of all their gas.
Thus we see that high angular momentum orbits in the outer parts of
the Virgo cluster could allow slow gas removal from accreted field
galaxies.

\subsubsection{Kelvin-Helmholtz Instabilities}

The interface between the Virgo intracluster medium and the interstellar 
medium of member galaxies produces a Kelvin-Helmholtz instability which could 
gradually extract gas from these galaxies, aiding ram-pressure in removing 
gas.  The lower-mass limit for Kelvin-Helmholtz
instabilities to be efficient is roughly 10$^{3}$ times higher than the 
ram-pressure stripping limit (Mori \& Burkert 2000).  It is however
unlikely that gas rich
LMCGs are being depleted gradually through this process.  
The mass loss rate induced by Kelvin-Helmholtz instabilities is 
given by Nulsen (1982) as

\begin{equation}
\dot{M} = \pi R^{2} \times n(r) \times v,
\end{equation}

\noindent where $R$ is the radius of the galaxy, $v$ is the 
velocity of the galaxy relative to the intracluster medium and n(r) is
the mass density of the intracluster medium.  Using typical
values (Table 3) of $R$~=~1~kpc and $v$~=~500~\kms, we find that $\dot{M} 
\sim 10^{3}$ \solm Myr$^{-1}$.  The time-scale for gas depletion of a typical 
LMCG via Kelvin-Helmholtz instabilities is then $\sim~10~-~15$ Gyr.  

If a small galaxy entered the Virgo cluster sometime in
the last few Gyr on a high angular momentum orbit, then it would not yet have
been stripped of all its gas, either through ram-pressure or 
Kelvin-Helmholtz instabilities. Both of these processes are
occurring, but they are slowly removing gas from galaxies on high angular 
momentum orbits in the outer parts of the Virgo cluster.  On 
the other hand, it appears that these galaxies would
have been completely stripped of all their gas if they had existed within the 
cluster on the order of a Hubble time, no matter what their orbits are.
The above arguments again suggest that these HI LMCGs are objects that must 
have recently entered the cluster in the last few Gyr, or have just been
accreted.  

\subsubsection{Gas Replenishment}

During the normal course of stellar evolution stars in galaxies will deposit
gas into the interstellar medium, thereby increasing the presence of HI.
Gas is returned to the ISM from mass loss in stars during the AGB and 
planetary nebula phases.  
The amounts returned vary, depending on the age and total mass of the stellar 
populations (e.g., Faber \& Gallagher 1976).  Following
the arguments in Faber \& Gallagher (1976) for giant ellipticals and
Gnedin et al. (2002) for the globular cluster $\omega$ Centauri,
we calculate that the mass returned to the ISM for a LMCG with a stellar
mass of $\sim 10^{8}$~\solm is $\sim 10^{6}$~\solm\,~Gyr$^{-1}$.  If we assume
that the HI LMCGs have been in the cluster for a few Gyr, then not
enough time has elapsed to produce the observed HI gas content of the
HI detected early-type LMCGs from evolved stars.  However, if this was
the complete story, we might expect substantial HI content to be a common 
feature of very old early-type LMCGs, but we see very little HI in 
field ellipticals galaxies or in low-mass cluster galaxies 
(e.g., Knapp, Kerr \& Bowers 1978; Young \& Lo 1997; Sage, Welch \& 
Mitchell 1998).   It is still not known with certainty
why giant elliptical galaxies are depleted of cold interstellar gas, but what
ever the cause
of depletion, it is likely occurring in LMCGs as well.  In summary, we do not
consider the return of gas from stellar evolution a likely scenario to
explain the HI gas content of these HI LMCGs.

\subsubsection{Impulsive Encounters - Dynamical Stripping}

The high number density and large relative velocities of cluster galaxies
makes impulsive encounters between them common, and
can result in mass loss, such as in the harassment scenario
(e.g., Moore et al. 1998), although this effect is likely not important
for the HI LMCGs since they are located in the outer parts of Virgo.   
The energy imparted to a galaxy with mass M$_{\rm gal}$ 
during an impulsive encounter between two spherical galaxies is proportional to
E $\alpha$ M$_{\rm pert}$M$_{\rm gal}$ b$^{-4}$V$^{-2}$ (Spitzer 1958) 
where M$_{\rm pert}$ is the 
mass of the perturbing galaxy, b is the distance between the two galaxies 
and V is the relative velocity between them. The large separation
of the HI LMCGs from the center of Virgo is an indication that the factor
b is much larger than for most cluster galaxies.  At a typical LMCG impact
parameter of
of 0.5 - 1 Mpc, the energy imparted to a galaxy and its resulting mass
loss due to impulsive encounters is very small (Aguilar \& White 1985). The 
strongest effects of mass loss involve cluster galaxies on
nearly radial orbits where the parameter b becomes small (Moore et al. 1998).
If the HI LMCGs are on high angular momentum orbits, as has been argued, 
then they do not feel the strong effects of impulsive encounters and thus 
will not lose a significant amount of mass from these interactions.  This is 
especially
true if the HI LMCGs have only been in the cluster for a short time.
We conclude therefore that the only processes strongly affecting the HI LMCGs
are ram pressure induced gas loss.

\vspace{0.1in}

\subsection{Morphological Transformations}

The question we address in this section is whether or not the HI detected
LMCGs have recently been accreted into Virgo (and not yet been stripped of
their gas), or if they have been in the cluster for at least a few Gyr, 
having evolved from a precursor of different morphology.  As we argued in 
\S 4.2 - 4.3, it is unlikely that these galaxies are an old cluster 
population.  Here we conclude that these LMCGs originate from accreted
field populations.

The idea that galaxies can undergo a morphological transformation in 
clusters has considerable observational support.   Distant
clusters contain large populations of blue, distorted galaxies (e.g.,
Oemler, Dressler, \& Butcher 1997; Couch et al. 1998), which are not
seen in the same proportion in nearby clusters (cf. Conselice \& Gallagher 
1998, 1999).   There is also evidence that S0 galaxies are less common in
moderate redshift clusters than in comparison to similar clusters at low 
redshifts (Dressler et al. 1997).  Furthermore, in Paper I we argue that
all galaxy types, except giant ellipticals, have kinematic and
spatial properties suggesting they were accreted into Virgo after
its initial formation, perhaps due to hierarchical accretion into the cluster 
(Kauffmann 1995).  Once they enter the cluster, galaxies can 
morphologically evolve into earlier 
types (Moore et al. 1998; Mao \& Mo 1998; Abadi et al. 1999; Quilis et al. 
2000).  What is missing from these arguments is whether or not galaxies in 
nearby clusters are undergoing evolution, or morphological transformations, 
as they should if any are still being accreted into clusters, albeit
at reduced rates compared to clusters at high redshift 
(e.g., Tully \& Shaya 1984). Previous
observations suggest that there are some small galaxies in the Virgo
cluster that appear to be undergoing rapid evolution (Gallagher \& Hunter 
1989; Vigroux et al. 1996), but the extent of this process
is currently unknown.    Does most active evolution of cluster
galaxies, including morphological transformations, occur only
in the distant past?    We argue here that HI LMCGs are possible 
examples of objects slowly undergoing transformations today.  The 
evidence, which we discuss in detail below, includes:  the outer 
spatial distribution of these detections in Virgo, their mixed morphological
appearances, their gas and stellar content, and the correlation between HI 
line-width and total B-band magnitude. 

\subsubsection{Morphologies and Stellar Populations: Dwarf Elliptical-Like}

We can try to understand when and how stars in HI detected LMCGs formed by 
examining their morphologies and the limited spectral information we have.
The morphologies and structural parameters of these galaxies (Figure~2-3, 
Table~4) are clearly not influenced by star formation.
The red colors of VCC~168 and VCC~608 demonstrate that these objects contain
older stellar populations, and the clumpiness parameter $S$ is low enough
that any star formation in these HI LMCGs must be smoothed out.  The
question is, how long does it take for a young stellar population to become
this red and smooth? A galaxy's stellar populations will be 
smoothed out after approximately a crossing time when young stars die and the
population dissolves into the background (Harris et al. 2001) which takes a 
few Gyr for these galaxies, especially if their random motions
are increased slightly through dynamical heating (Moore et al. 1998).  
For a stellar population to become 
as red as VCC~168 and VCC~608 from a blue star-forming irregular, or disk 
galaxy, requires roughly the same amount of time.   These
HI detected LMCGs therefore have stellar population and morphological 
characteristics of dwarf ellipticals, and must have had their last major 
episode of star formation several Gyr ago.
\newpage
\subsubsection{Gas Content: Dwarf Irregular-Like}

A diagnostic for understanding the origin of HI-detected
LMCGs is to compare their optical and HI properties,
including M$_{\rm HI}$/L$_{\rm B}$ values, absolute magnitudes and
HI masses, to those of other galaxy populations, including the Gallagher \& 
Hunter (1989) Virgo cluster dwarf irregular and amorphous samples.  
Table 5 lists the averages of several of these quantities as a function of
galaxy type.   The HI gas
mass fractions and M$_{\rm HI}$/L$_{\rm B}$ ratios are fairly similar in
the various galaxy populations, except for the Local Group dEs,
where these values are significantly lower.
Dwarf irregulars in the Local Group have the highest f$_{\rm gas}$ and
M$_{\rm HI}$/L$_{\rm B}$ ratios, while the Local Group dEs have the lowest.   

For example, the gas mass fraction, f$_{\rm gas}$, for dwarf irregulars
in the Local Group are 0.07, 0.08, 0.2, 0.65 and 0.46 for the Large
Magellanic Cloud, NGC 6822, IC 10, IC 1613 and DDO 221. The
average f$_{\rm gas}$ value for all Local Group dwarf irregulars 
is 0.30$\pm$0.24 (Table 5).    Virgo LMCGs classified as dEs with
HI detections in Virgo have an average f$_{\rm gas}$ value of 0.30$\pm$0.28
(Table 3). On the other hand, the gas mass fractions for Local Group dwarf 
spheroidals are often
$<$~10$^{-4}$~-~10$^{-5} \sim 0$ (Johnson \& Gottesman 1983; 
Koribalski et al. 1994; Young \& Lo 1997; Young 1999). 
  
\begin{inlinefigure}
\begin{center}
\resizebox{\textwidth}{!}{\includegraphics{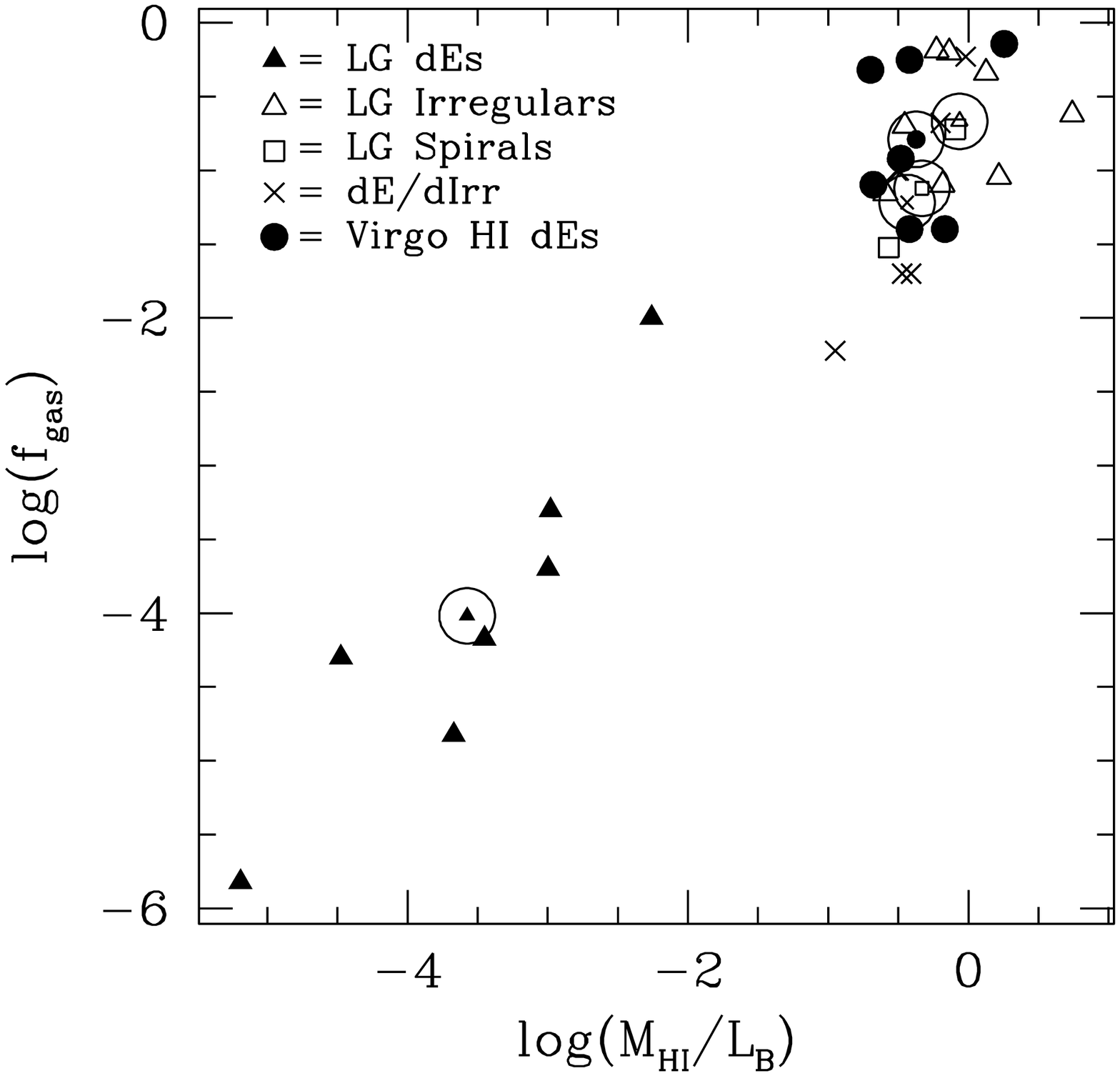}}
\end{center}
\figcaption{The fraction of mass in HI (f$_{\rm gas}$) for various Virgo 
cluster member galaxy types
plotted as function of M$_{\rm HI}$/L$_{\rm B}$.  Each separate galaxy 
population is designated by the labeled symbols.  The average position for 
each type is plotted as a small version of its respective symbol in a circle.}
\end{inlinefigure}

Some M$_{\rm HI}$/L$_{\rm B}$ values for the HI detected LMCGs are as high or 
higher than some spirals and irregulars, although some gas rich
low-luminosity early types have been found (Sadler et al. 2000).  Even if 
some gas mass is lost through
various stripping effects, the fading of stellar populations due to
the truncation of star formation (\S 4.4.4) may be able to reproduce
this.  Stellar populations with initial colors similar to late-type spirals and
irregulars will become $\sim 1.5$ magnitudes fainter in $\sim 3$ Gyr, easily
allowing for high M$_{\rm HI}$/L$_{\rm B}$ values even if some gas is
removed.  Some of these systems could also be low M$_{\rm HI}$ accreted
field galaxies, as some examples are known to exist in the field 
(Knezek, Sembach \& Gallagher 1999).

Average values of other properties, listed in Table~5, show that Local Group 
dwarf ellipticals/spheroidals have very different derived HI gas properties 
from the Virgo HI detected LMCGs classified as dEs.  In 
fact, the average values are different at a 7$\sigma$ significance.
Based on this, the HI dEs cannot simply be recently 
accreted analogs of Local Group dwarf ellipticals/spheroidals, as they have
too much HI.   Although the HI detected LMCGs studied here morphologically 
appear to be dwarf ellipticals, they could not have originated
from dEs in groups, as no pure dwarf elliptical has a gas content as large
as these LMCGs.  We propose that the only way these galaxies could
contain such a high HI gas mass is for them to have originally been  
star-forming systems that morphologically evolved into a dE-like object over 
the last few Gyr through slow gas removal and passive evolution.  
The most similar galaxy type to the Virgo HI detected LMCGs are the Local 
Group transition types (van den Bergh 2000), which include Phoenix and 
Pisces, and irregulars, which have f$_{\rm gas}$ and M$_{\rm HI}$/L$_{\rm B}$ 
values within 1$\sigma$ of the LMCG values (Figure~10), although the HI 
masses for these LMCGs are more similar to the values found for the 
irregulars.  As we argue in \S 4.4.4, star formation in 
these systems could be suppressed.

\subsubsection{Parameter Space: Transition Types}

We can further use diagnostic plots such as Figures~10 through 13 to 
understand which of the well-understood nearby galaxy populations is most 
similar to the HI LMCGs.
Figure~10 plots f$_{\rm gas}$ and M$_{\rm HI}$/L$_{\rm B}$ values for 
individual Local Group spirals 
(open boxes), irregulars (open triangles), dEs (solid triangles), transition
types (crosses) and the Virgo HI LMCGs studied in this paper 
(solid circles).  The average f$_{\rm gas}$ and M$_{\rm HI}$/L$_{\rm B}$ values
for each type are labeled by a circled small version of their respective
identification symbol, and the average values are listed in Table~5.
Again, from this figure it appears that the HI detected LMCGs are most
similar to the Local Group transition types in terms of their HI gas and 
stellar properties.

We can also examine the differences between the Virgo dEs with HI compared
to Virgo cluster dwarf irregulars observed by Gallagher \& Hunter (1986, 
1989). Figure~11 plots M$_{\rm HI}$/L$_{\rm B}$ ratios vs. M$_{\rm B}$ for the 
HI LMCGs, nearby dwarf transition types, and the dwarf irregular and
amorphous galaxies studied by Gallagher \& Hunter (1989), converted to our 
assumed distance of 18 Mpc.  Figure~11 includes the Virgo amorphous galaxies, 
objects that morphologically
appear to be early-types but have evidence for star formation (Sandage
\& Brucato 1979).  Figure~11 shows that the Virgo LMCGs classified as dEs
with HI detections (solid circles) on average have M$_{\rm HI}$/L$_{\rm B}$ 
values between the values found for the gas rich Virgo irregulars and the 
amorphous galaxies, which are gas depleted (see Table 5).  
The nearby dwarf transition types in the Local Group are plotted as crosses 
on Figure~11 and their M$_{\rm HI}$/L$_{\rm B}$ distribution is very similar 
to the M$_{\rm HI}$/L$_{\rm B}$ ratios for HI LMCGs classified as dEs.  

\newpage

\subsubsection{The Four LMCG Populations (Yet Known)}

We can gain further insight into LMCGs by examining the limited information
we have on the stellar and gaseous properties of all the different types.
Figure~12 shows the correlation between the $(U-B)$ colors and the 
M$_{\rm HI}$/L$_{\rm B}$ ratios of the Virgo dwarf
irregular and amorphous samples from Gallagher \& Hunter (1989).  
As Figure~12 shows, the more gas-depleted
objects, as measured by the M$_{\rm HI}$/L$_{\rm B}$ ratio, have redder $(U-B)$
colors, indicating the increasing dominance of older and possibly
more metal rich stellar populations.   We also plot on Figure~12 the location 
of the two HI LMCGs with $(B-V)$ color measurements, VCC~168 and VCC~608.
We convert the $(B-V)$ colors for these objects into a $(U-B)$ color by
fitting, and using, the relationship between these two colors for all galaxies
in the RC3: $(U-B)~=~1.26\pm0.01~(B-V)~-~0.71\pm0.01$.   The HI detected
LMCGs have M$_{\rm HI}$/L$_{\rm B}$ values similar to Virgo dwarf irregulars, 
but have red colors, similar to the amorphous galaxies.  It appears
that these HI LMCGs might be members of a new class of LMCGs, 
pending proper $(U-B)$ color measurements. An arrow on Figure~12
shows where dEs with no detectable HI would fall on this diagram, as they are 
red and gas depleted systems.

\begin{inlinefigure}
\begin{center}
\resizebox{\textwidth}{!}{\includegraphics{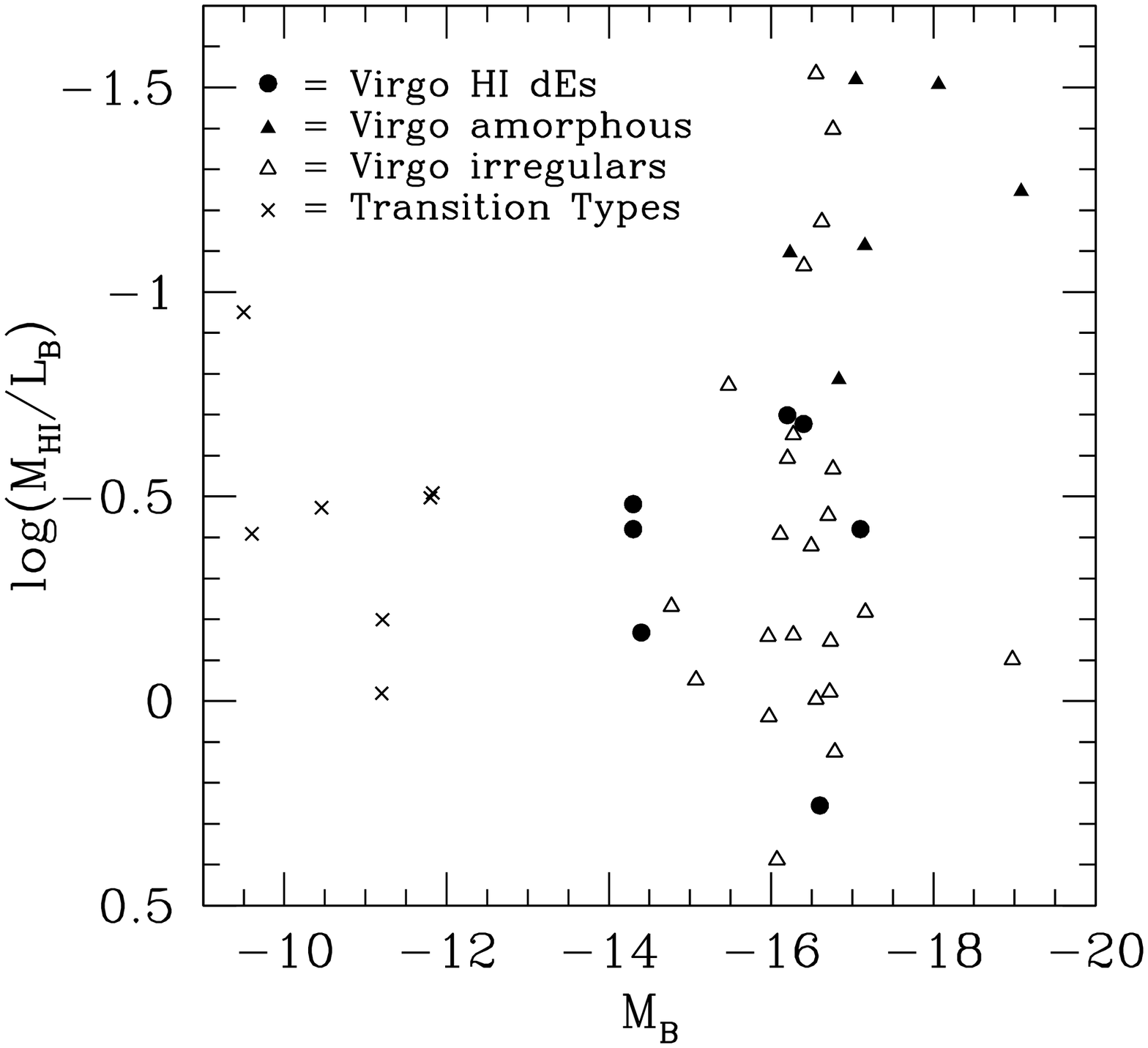}}
\end{center}
\figcaption{The absolute magnitude (M$_{\rm B}$) vs. M$_{\rm HI}$/L$_{\rm B}$ 
diagram for
various galaxy populations, including Virgo irregular and amorphous
galaxies from Gallagher \& Hunter (1989), the dE HI sample studied 
in this paper, and dwarf transition types in the Local Group.}
\end{inlinefigure}

Based on this there appear to be
at least four types of low-mass galaxies associated with the Virgo cluster: 
1. the classical dwarf irregulars with active star formation and gas, which 
are likely in the process of being accreted into
clusters (Gallagher \& Hunter 1989; Paper I); 2. dwarf ellipticals
with older stellar populations and HI gas masses $< 10^{6}$ \solm that
were accreted $> 3$ Gyr ago, some as larger
mass galaxies, or those that originally formed with the cluster (Paper III);  
3. amorphous galaxies; and 4. the HI rich early-type 
LMCGs studied in this paper.  These last two types are likely LMCGs undergoing
morphological evolution from accreted low-luminosity disks or irregulars.  
The amorphous 
galaxies, because of their low HI content, and red colors, are possibly systems
that were accreted into eccentric orbits, or under physical conditions such 
as within groups, where they lost a significant amount of their HI gas 
mass.   HI LMCGs, discussed in this paper, are potentially dEs which may have 
retained their HI gas because they were accreted into the cluster
on high angular momentum orbits.  The redder colors of VCC~168 and VCC~608, 
and the morphologies of the HI LMCGs, both suggest that HI LMCGs have existed
in the cluster for a longer time than the amorphous galaxies. 

\begin{inlinefigure}
\begin{center}
\resizebox{\textwidth}{!}{\includegraphics{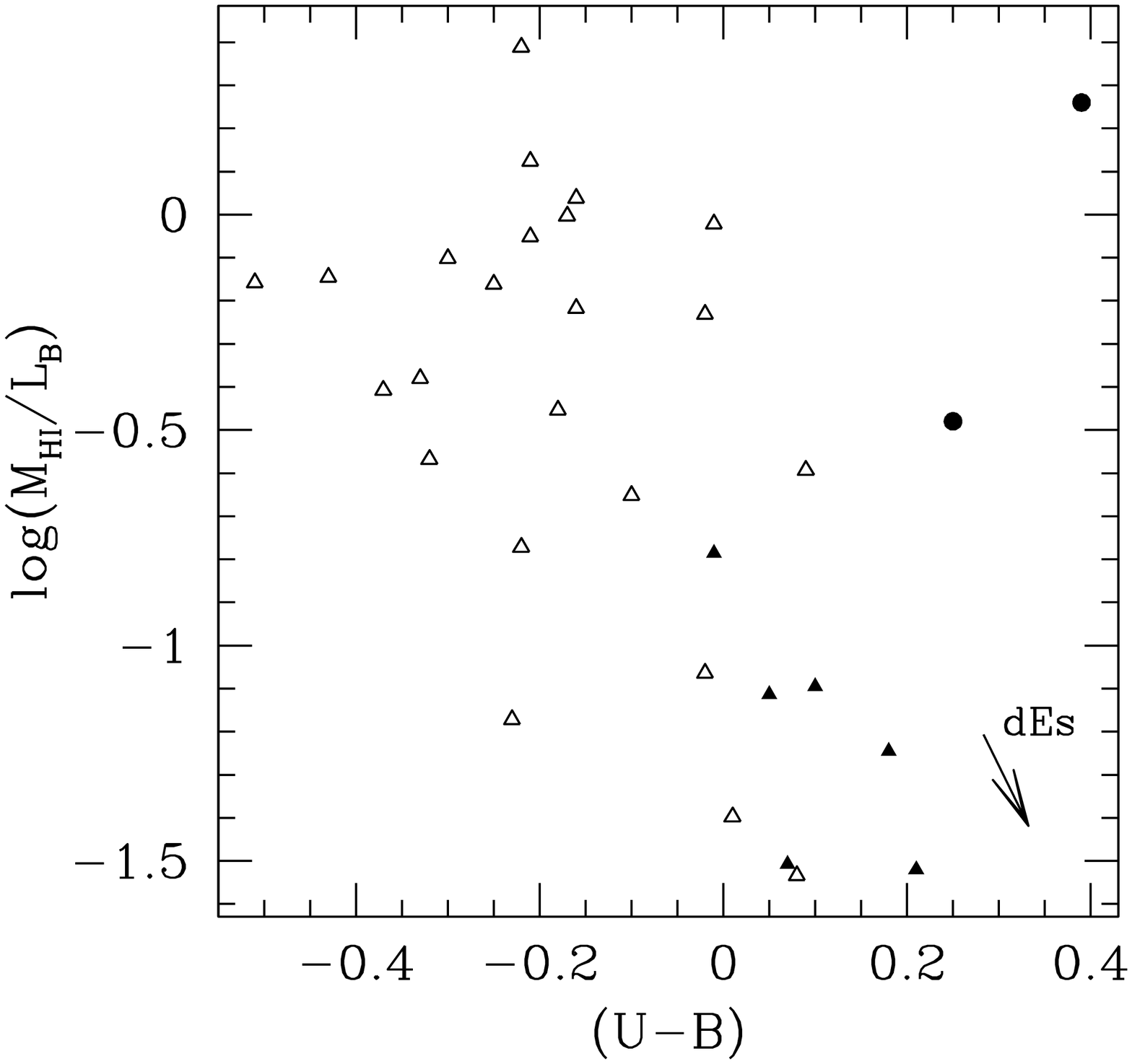}}
\end{center}
\figcaption{Diagram of M$_{\rm HI}$/L$_{\rm B}$ vs. $(U-B)$ color for
the irregular (open triangle) and amorphous (solid triangle) Virgo galaxy 
sample from Gallagher \& Hunter (1986).  The two HI detected LMCGs with both
M$_{\rm HI}$/L$_{\rm B}$ and derived $(U-B)$ colors, VCC~168 and
VCC~608, are plotted as solid circles.}
\end{inlinefigure}

In the above picture we are not arguing that all dwarf ellipticals 
originate from dwarf irregulars, as the usual objections for this
evolutionary process still hold (Paper I).  The only dwarf 
ellipticals/spheroids
that originate from the irregulars would be those that are faint with
low surface brightnesses.  The origin of the brighter, higher surface
brightness objects is potentially dynamical in nature (Paper I and III,
Conselice 2002).

\subsection{The Luminosity-Line Width Relationship}

Figure~13 shows the luminosity line width diagram for the HI LMCGs detections 
listed in Table~3, minus the two galaxies we
conclude are either not likely a dE or a false HI detection.
With the exception of VCC 390, which has a high
\dml ratio of $\sim$45, there is a general correlation between the absolute
magnitude M$_{\rm B}$ and the velocity width W$_{20}$.   Also
plotted on Figure~13 are the Faber-Jackson and Tully-Fisher relationships
between M$_{\rm B}$ and the velocity line width, or internal
stellar velocity dispersion.  Data from
which these relationships are fitted are shown, and are from 
Faber \& Jackson (1976) and Pierce \& Tully (1992)\footnote{We convert the 
magnitudes in Faber \& Jackson's (1976)
photometry to match the Hubble constant, 85 km s$^{-1}$ Mpc$^{-1}$ used
in Pierce \& Tully (1992).   The magnitudes
for the spiral galaxies in Pierce \& Tully (1992) were 
corrected for internal and galactic extinction.  The velocity widths, 
W$_{\rm R}$, plotted for the spiral data are corrected using the methods 
outlined in Tully \& Fouqu\'e (1985).}.    Also plotted in Figure~13, as 
crosses,
are very late-type disk galaxies, which are good candidates for
being LMCG progenitors, taken from Matthews et al. (1998).

\begin{inlinefigure}
\begin{center}
\resizebox{\textwidth}{!}{\includegraphics{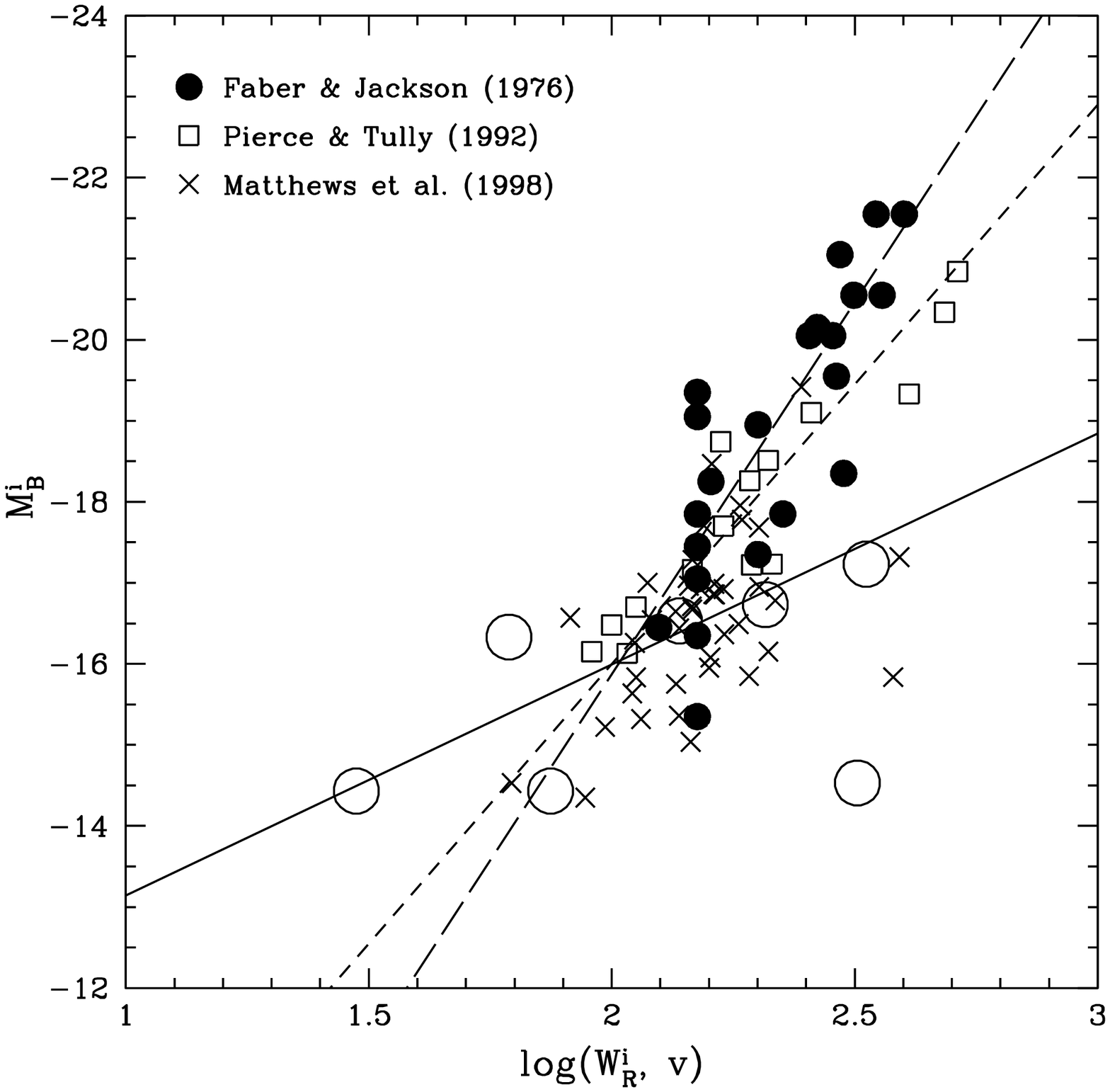}}
\end{center}
\figcaption{The velocity line-width (W$_{\rm R}^{\rm i}$) or velocity 
dispersion (v) vs. luminosity (M$_{\rm B}^{\rm i}$) 
relationship for various types of galaxies.  The solid circles are the 
galaxies from the sample in Faber \& Jackson (1976) and the open boxes are 
from the sample used in Pierce \& Tully (1992).  The long dashed line is a 
fit to the Faber-Jackson points, while the short dashed line is the 
Tully-Fisher relationship and the crosses are for the extreme late-type 
disk galaxies from Matthews et al. (1998).
The open circles are the LMCGs after correcting their velocities for
inclination and removing in quadrature a minimum value for turbulence, 
W$_{\rm t} = 5$ km s$^{-1}$.}
\end{inlinefigure}

The magnitude and velocity data for the HI LMCGs listed Table~3 are plotted 
as open circles, and represent the magnitudes and W$_{20}$ values after 
correcting for inclination and subtracting out a component to account for 
turbulence.  While the turbulent velocity contribution
could be fairly high in some galaxies (Tully \& Fouqu\'e 1985;  de 
Blok, McGaugh \& van der Hulst 1996),
it is likely lower in dwarfs, $\sim$5 \kms\, based on measurements of
nearby low-mass galaxies (e.g., Young
\& Lo 1997).  We therefore remove 5~\kms\, from the W$_{20}$ velocity
widths when correcting for turbulence.

The solid line shows a best fit relationship between the galactic extinction
corrected M$_{\rm B}$ values and velocity widths for the HI LMCGs.
This fit shows that the LMCGs classified as dEs with HI 
gas appear to deviate from
the Tully-Fisher and Faber-Jackson relationships in both slope and zero point,
and are more similar to the trends, shallower slope, and individual values 
found for the extreme late-types in the Matthews et al. (1998) sample.
This relationship is also fairly flat and several galaxies clearly deviate 
from it.  Figure~13 
shows that the galaxy sample under consideration, while displaying a 
luminosity line-width correlation on average (solid line), has a 
fundamentally different nature than the relationship for both the spirals and 
ellipticals.   While e.g., Matthews et al. (1998) find that some faint
galaxies have high velocities for their magnitudes compared to Tully-Fisher, 
revealing a possible high
mass to light ratio, we do not generally see this trend for all the HI
detected LMCGs.  

There are several possible ways to account for the dispersion in the 
relationship between HI line width and absolute
magnitude in the HI detected early-type LMCGs, besides observational error.  
One is 
that these galaxies are in a mix of evolutionary stages, such that the velocity
width is sensitive to not only the mass of the galaxy, but also whether or not
the gas disk has been extended or truncated by cluster processes such
as gas and tidal stripping (Moore et al. 1998).  Models of the 
evolution of low-mass galaxies in groups also show that as a galaxy interacts 
with a larger system and losses mass, its mass to light ratio 
varies with time depending on whether any new star formation occurs, and what 
the mass profiles of the various components are (e.g., Mayer et al. 2001).
The mass to light ratio could also be somewhat overestimated by the material 
stripped
from galaxies that can produce a larger observed velocity dispersion, and a
range of derived M/L ratios (Mayer et al. 2001).  This would however
require non-virial motions of the gas, that are unlikely if ram-pressure
forces are reduced in intensity (\S 4.3).  

\section{The Recent Virgo Accretion Rate}

We can use the information on the number of transition type
dwarfs in the Virgo cluster to get an idea of the recent galaxy
accretion rate which might be responsible for producing these 
objects.    To obtain this number we follow the procedure presented by 
us in Gallagher et al. (2001).

Using Gallagher et al. (2001) and assuming a cluster age
of 10~Gyr, and a dE production efficiency from all but giant galaxies of
100\%, then since there are $\sim$10$^3$ dEs in Virgo with 
M$_{\rm B} \leq -14$, this requires an 
average infall rate of $\geq$100 galaxies per Gyr.  High infall rates in
the past are therefore necessary to populate modern clusters with 
low-mass cluster objects, but higher past cluster infall rates are likely 
in a low $\Omega_{\rm M}$ universe (Kauffmann 1995).

Infall rates are difficult to measure, and probably will be episodic,
since galaxies are observed to accrete into Virgo in groups 
(e.g., Tully \& Shaya
1984).  One estimate can be derived from the Gallagher \& Hunter (1989) sample
which suggests that
a few dozen galaxies are currently undergoing morphological transformations. 
This is similar to the $\sim 20$ galaxies estimated to have
recently fallen into Virgo in the last Gyr (Tully \& Shaya 1984).  We obtain 
another estimate of the total
number of current transition dwarf objects in the Virgo cluster using
the results of this paper. A total
of 49 objects classified as dwarf ellipticals were observed for HI for
which a total of 7 are likely transition objects.  If we
assume that these 49 are representative of the entire Virgo dwarf
elliptical population, and ignore any transition types that may
be classified as dwarf irregulars, then $\sim$15\% of all dwarf ellipticals
with M$_{\rm B} < -14$
in the Virgo cluster are in some kind of transition phase.   As there
are roughly 10$^3$ dwarf ellipticals in the Virgo cluster (Binggeli
et al. 1985), then we expect $\sim$160 of these galaxies to be in
transition throughout 
Virgo.  If we assume that an average transition phase takes
1-3 Gyr (e.g., Moore et al. 1998) then the recent
galaxy infall rate in the last 3 Gyr (from z$\sim 0.2$ in $\Lambda$CDM) is 
$\dot N_c(t_{0})$ $\approx$ 160 galaxies / 3 Gyr $\sim$ 50 galaxies/Gyr.  
This is an upper bound as some mixed structure galaxies could
be on nearly circular orbits, and thus slowly evolving. 

We can use further assumptions about the
average amount of mass per galaxy, to derive a mass infall rate of
dwarfs into the Virgo cluster in the last 3 Gyr.  The average absolute
magnitude for the HI detected LMCGs is M$_{\rm B} = -15.6\pm1.2$.
This gives an average mass per galaxy, assuming a stellar M/L ratio of
5 in solar units, of $\sim$$5 \times 10^8$ \solm.  Using this approximation
for our hypothetical 160 transition dwarfs accreted over the last 3 Gyr 
gives a mass influx into the Virgo cluster of $\sim$ 50 \solm/yr only
from this low-mass component.  Assuming a normal luminosity function for
infalling galaxies the total time average infall rate would be several
times larger than this, $\sim 200 - 300$ \solm/yr.  
While we have used a very simplistic model to 
obtain this result, it shows that an accretion model for HI LMCGs  
evolving into dwarf ellipticals gives reasonable galaxy and mass
infall rates, comparable to other Virgo cluster accretion
estimates (e.g., Tully \& Shaya 1984). 

Are these calculations of the recent and past accretion history into clusters
consistent with cosmological model predictions?  In a $\Lambda$-dominated 
flat cosmology, the linear growth of structure slows
down at intermediate redshift, $1+z \sim \Omega_m^{-1/3}$ 
(note this is a lower redshift than that at which linear growth slows for a
matter-dominated open universe with the same $\Omega_m$), but growth is
a process that continues until the present day, with larger structures
assembling a larger fraction of their mass later. Simulations designed
to constrain cosmological parameters through the analysis of the
amplitude of substructure in clusters (Crone, Evrard \&
Richstone 1996) have shown that in a model
with $\Omega_\Lambda = 0.8$, $\Omega_m = 0.2$ and a power spectrum
similar to that of CDM over cluster mass scales ($n =-1$), 50\% of the
mass of a typical large cluster was assembled within the last
5~$h^{-1}$ Gyr.  The $\Lambda$CDM simulations of Wechsler et al.~(2002)
also show continuous accretion and mass assembly to the present day,
with their most massive haloes, $M > 3 \times 10^{13}h^{-1}M_\odot$,
somewhat lower than a typical galaxy cluster, accreting around 25\% of their
mass since $z \sim 0.7$ (their Fig.~4a), or a look-back time of $\sim
7$~Gyr.  These haloes accrete typically about 10\% of their mass in
the last $\sim 3~$Gyr ($z \sim 0.25$).  The `Virgo-like' clusters
simulated by Governato, Ghigna and Moore (2001) also accrete $\sim 20$\% of
their mass since $z=0.25$ in the `concordance' $\Lambda$CDM universe
(see their Fig.~5; a similar recent accretion history is found in the
open CDM model).  These CDM models are therefore consistent with the idea, 
that Virgo HI rich early-type LMCGs are a recently accreted galaxy population.

\section{Summary and Conclusions}

In this paper we present evidence that HI-rich Virgo cluster dEs, discovered 
through an Arecibo 21-cm and WIYN 3.5-m telescope study, form 
a new class of low-mass galaxies. 

\begin{enumerate}

\item We find two new low-mass galaxy HI 21-cm line detections with the 
Arecibo telescope in the Virgo cluster, out of 22 observed objects
classified as dEs with optically determined radial velocities. These
dwarfs, VCC~390 and VCC~1713, have M$_{\rm HI}$$= 6 \times 10^7$ \solm and $8
\times 10^7$ \solm, respectively.  The other 20 Virgo dE members we observed
were not detected, with 3$\sigma$ upper limits of M$_{\rm HI}$$\le 8 \times 
10^6$ \solm, corresponding to M$_{\rm HI}$/L$_{\rm B} < $0.1 (see Table 1). 

\item A survey of the
literature yields an additional 5 Virgo early-type dwarfs with credible HI
measurements, for a total of 7 detections among 49 galaxies with sensitive
21-cm line observations (Table 2). About half of all Virgo dEs
with measured radial velocities (see Paper I) now have been
searched for in HI to the $\sim 10^7$ M$_{\odot}$ level with a 15\% HI 
detection rate.

\item Optical imaging with the WIYN telescope for 5 of these Virgo 
early-type LMCGs with 21-cm detections allowed us to perform morphological 
classifications
using quantitative techniques. One of our new detections, VCC~1713, shows
clumpy sub-structure, which could be due to internal dust or residual
effects of star formation; it appears to be in a relatively early stage of
transition from a star forming galaxy to a dE like system. We further 
suspect that VCC~2062 is a dI galaxy, as it has a
faint, patchy structure on DSS images, and we removed it from the early-type
Virgo LMCG sample. The morphologies and other known optical structural
characteristics of the remaining 5 HI-detected galaxies are consistent
with those of normal dE members of the Virgo cluster. The optical
spectrum of one of our detections, VCC~390 (Figure~4), is also very
similar to other known dE spectra. The Virgo HI-rich
LMCGs have gas masses well above the levels found in
Local Group dEs, e.g., NGC~185 and NGC~205.  These relatively gas-rich 
galaxies may constitute $\sim$15\% of 
Virgo's moderate luminosity population of early-type dwarf galaxies.

\item The early-type LMCGs in Virgo with HI are found preferentially
near the edge of the cluster, at radii $\ge$ 0.5~Mpc from the cluster center, 
and have a flat distribution of observed radial velocities, suggesting that
they have recently been accreted (and not crossed the cluster yet), or 
are on high angular momentum orbits and are possibly accreted in groups.

\item We show that the HI LMCGs studied in this paper have 
HI properties more similar to Local Group transition type dwarfs 
or dwarf irregulars, rather than dEs or Virgo amorphous galaxies.

\end{enumerate}

\noindent We use this observational information, along with models of 
various cluster driven galaxy evolutionary processes, especially gas removal
due to ram-pressure 
stripping and Kelvin-Helmholtz instabilities, to conclude the following about 
these HI LMCGs and other low mass cluster galaxies in Virgo.

\begin{enumerate}

\item  Based on the positions of HI LMCGs in the outskirts of Virgo and their
high velocities relative to the cluster, we conclude that some, or all, of the 
HI-rich Virgo dEs are on high angular momentum orbits, and therefore 
never go through the cluster core.  This is a natural
explanation for how these galaxies are able to keep their HI gas over
long time spans.  If these galaxies had highly eccentric orbits, and passed 
through the Virgo center,
they would be rapidly stripped of their gas in less than a cluster crossing 
time.  Calculations of gas loss due to ram-pressure and 
Kelvin-Helmholtz instabilities show that these galaxies would be depleted of 
HI gas, even at their projected position, over $\sim$ 10 Gyr, but can 
potentially keep their gas when on high angular momentum orbits for up to 
a few Gyr.  

\item Through a comparison of gas properties of Local Group dwarf ellipticals
and spheroidals, which have much lower gas content than the Virgo HI-rich
LMCGs classified as dEs, we conclude that these galaxies must have
been of a different morphological type in the past.  This is 
consistent with the idea that these galaxies were accreted as gas rich 
star-forming systems into Virgo during the 
last 1 - 3~Gyr, as this is approximately the same amount of time needed
to convert a disk or irregular galaxy with star formation 
into a red early-type dwarf.   Even at the present epoch,
galaxy clusters act as engines of evolutionary change. They effectively
strip gas and stars from later-type low mass galaxies captured from the
cluster surroundings, and
thereby could convert low luminosity disk galaxies and irregulars into 
early-type cluster dwarfs.  We also find two very high HI line widths that
need to be explained.

\item Combining this analysis with the dwarf irregular and amorphous
galaxy properties from Gallagher \& Hunter (1989), and Paper I, we can sketch
a possible evolutionary/formation scenario for low-mass cluster
galaxies in Virgo.  The dwarf irregulars are recent additions into clusters
that have not yet crossed through the cluster, but when they do they will be 
stripped of much of their gas.    The presence of HI in some Virgo
LMCGs classified as dEs is additional evidence that some LMCGs are
created from infalling field galaxies (Paper I \& III).  Other dwarf 
ellipticals in Virgo have likely been present since the cluster was formed.  
We argue that the amorphous and HI LMCGs are both accreted populations,
either former disk galaxies or dwarf irregulars, whose evolution has
diverged due to their orbits in the cluster.  The amorphous galaxies are 
possibly on radial
orbits, or systems in previous collisions (Gallagher \& Hunter 1989), whose 
HI gas has been stripped, and are now passively evolving into
redder systems.  The HI LMCGs are, in our scenario, systems on circular 
orbits whose HI gas remains, and whose stellar populations and morphologies 
evolved to resemble dwarf ellipticals.  In this model, the HI LMCGs were 
accreted before the amorphous galaxies to allow time for morphological 
evolution.

\item  A simple calculation of expected Virgo cluster infall rates based on 
the number of low-mass galaxies observed in
evolutionary transition phases yields reasonable results, with a derived 
average recent mass infall rate of $\dot{\rm M}$ $\sim$ 
50 \solm\, year$^{-1}$, and
most likely a time averaged rate of 200 - 300 \solm\, year$^{-1}$, when
giant galaxies are included.

\end{enumerate}

We thank Chris Salter for his invaluable assistance in obtaining the data 
presented in this paper, and for his generous hospitality while CJC was
observing at Arecibo.  We also
thank Lynn Matthews for help taking the WIYN images, and for supplying 
us with velocities and magnitudes of her low-mass disk galaxy sample.  The
anonymous referee made many valuable points after a careful and thorough
reading of this paper.  This research was supported
in part by the National Science Foundation (NSF) through grants AST-9803018
to the University of Wisconsin-Madison and AST-9804706 to Johns Hopkins
University. CJC acknowledges support from a NSF Astronomy \& Astrophysics
Postdoctoral Fellowship,  a
Graduate Student Researchers Program (GSRP) Fellowship from NASA and the
Graduate Student Program at the Space Telescope Science Institute.

\begin{figure}
\plotfiddle{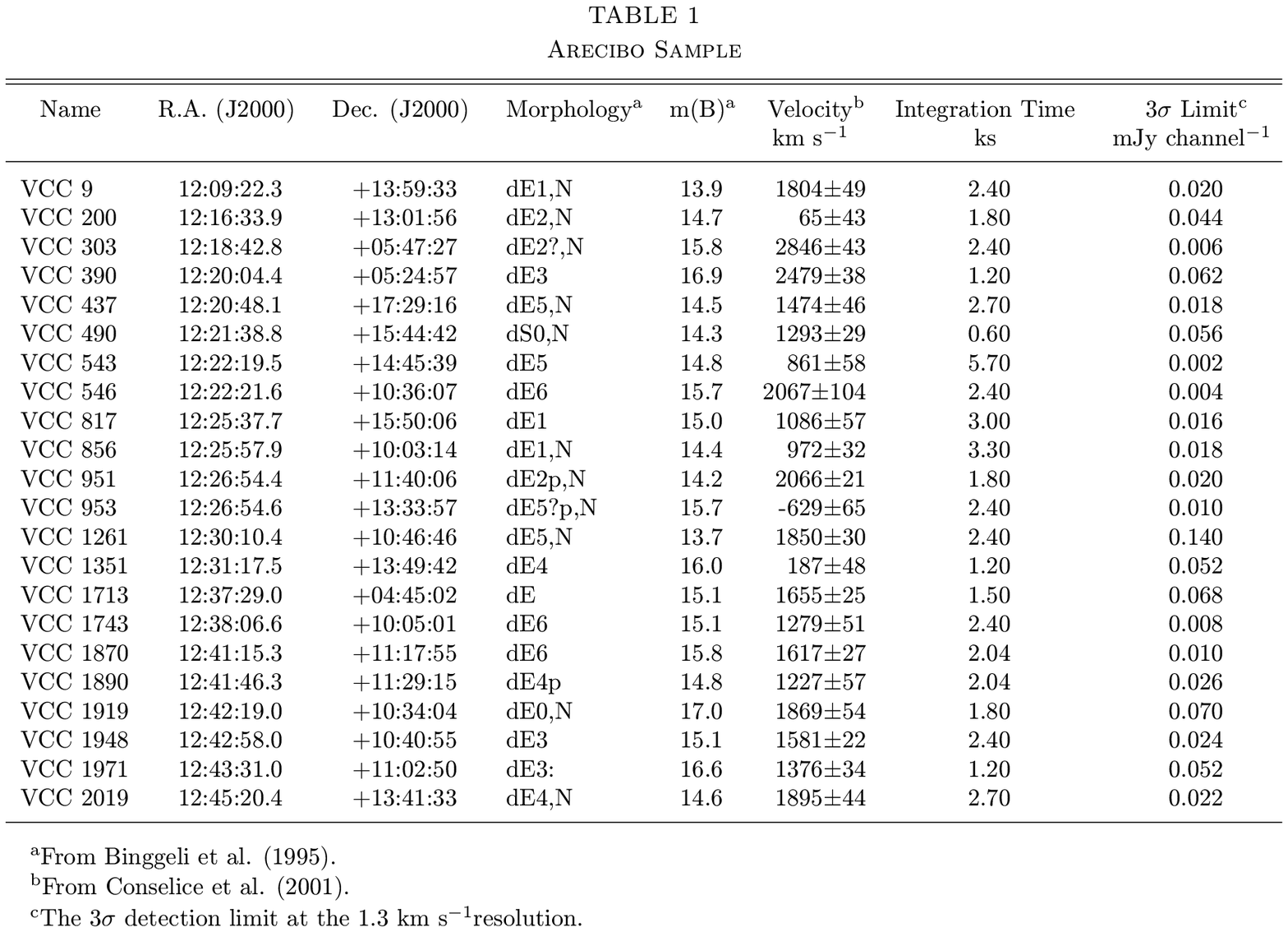}{6.0in}{0}{100}{100}{-310}{-170}
\end{figure}
 
\begin{figure}
\plotfiddle{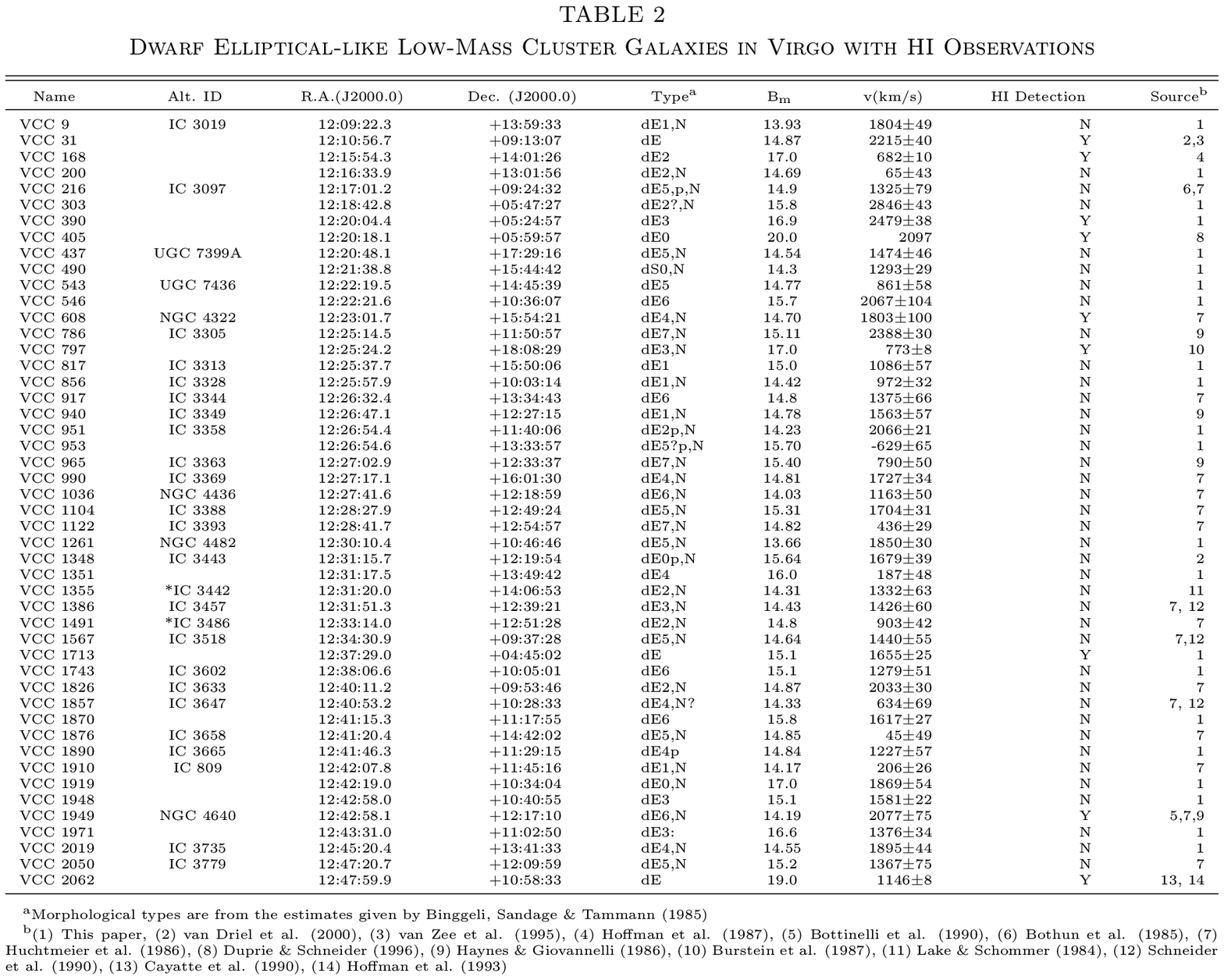}{6.0in}{0}{100}{100}{-310}{-170}
\end{figure}
 
\begin{figure}
\plotfiddle{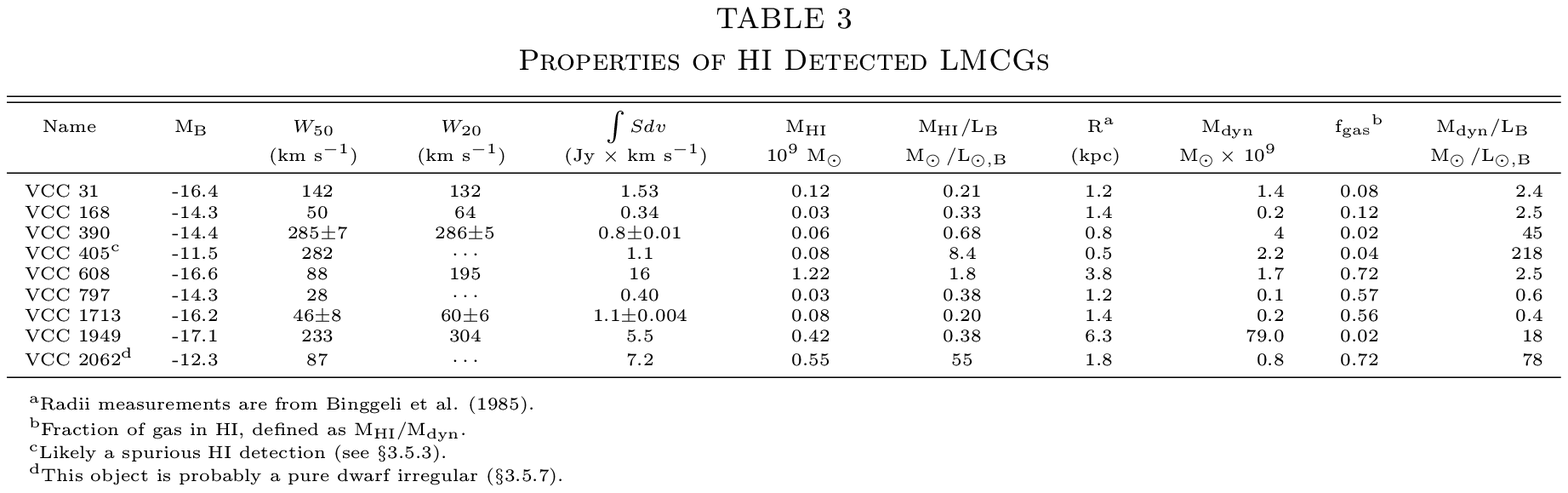}{6.0in}{0}{100}{100}{-310}{-170}
\end{figure}

\begin{figure}
\plotfiddle{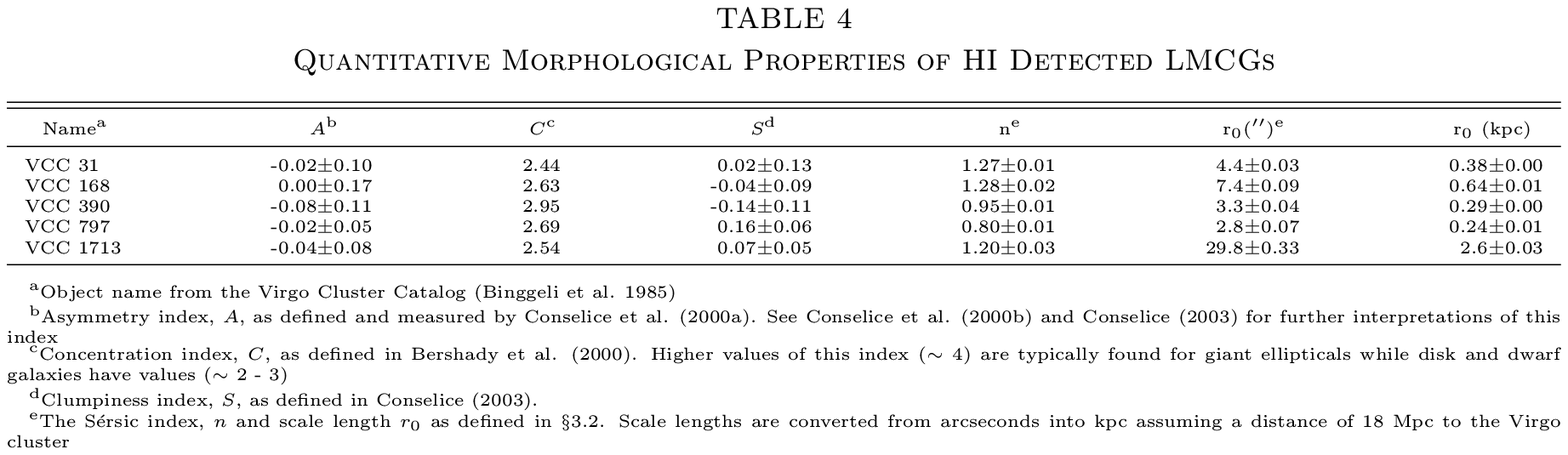}{6.0in}{0}{100}{100}{-310}{-170}
\end{figure}

\begin{figure}
\plotfiddle{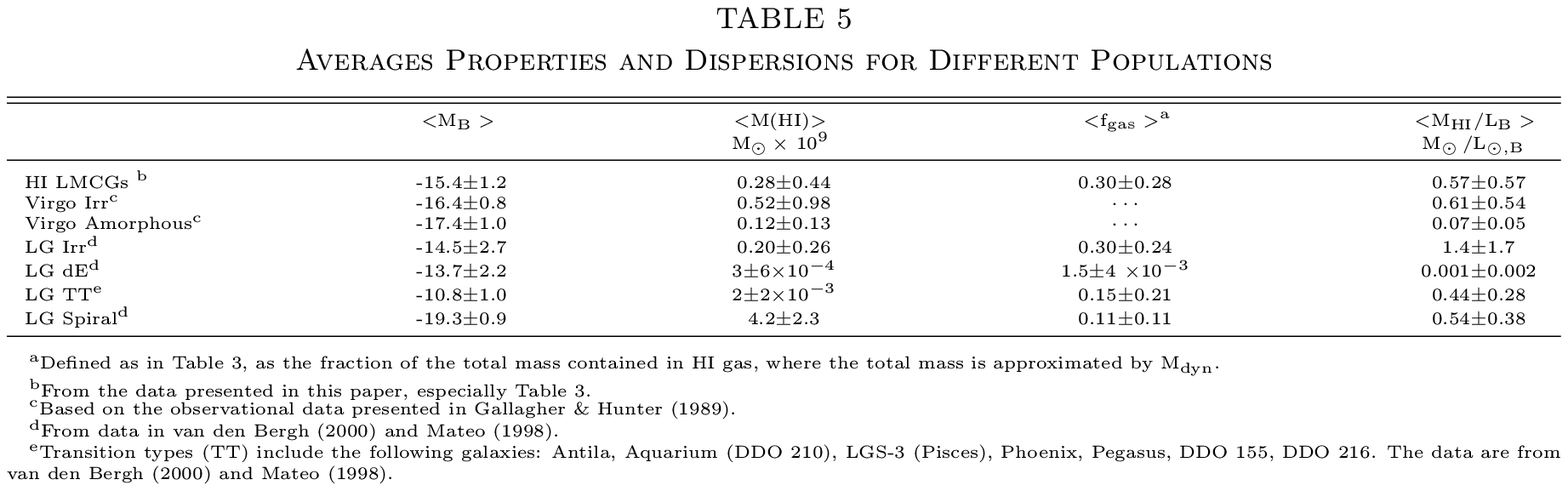}{6.0in}{0}{100}{100}{-310}{-170}
\end{figure}

\end{document}